\documentclass[aps,pra,twocolumn,superscriptaddress,showpacs]{revtex4-1}

\usepackage[utf8]{inputenc} % Required for non-English characters (input)
\usepackage[T1]{fontenc} % Required for non-English characters (output)
\usepackage[english]{babel}
\usepackage[autostyle=true]{csquotes} % Generate language-dependent quotes in the bibliography
\usepackage[dvipsnames]{xcolor}
\usepackage{graphicx}% Include figure files
\usepackage{array}
\usepackage{amsmath} % math formulas
\usepackage{amssymb} % math symbols
\usepackage{amsfonts} % math basic fonts
\usepackage{amsbsy}
\usepackage{mathrsfs} % math RSFS fonts (e.g. \mathscr)
\usepackage{dsfont} % math double strike (e.g. integer set, unity matrix with \mathbb)
\usepackage{dcolumn} % Align table columns on decimal point
\usepackage{upgreek} % Upgreek symbols (e.g. \Upbeta)
\usepackage{bm} % bold math
\usepackage{bbm}
\usepackage{placeins} % Control floats (e.g. \FloatBarrier)
\usepackage{physics} % Include bra and ket

\usepackage[colorlinks,linkcolor=blue,citecolor=blue,urlcolor=blue]{hyperref}
\usepackage[most]{tcolorbox}
\usepackage{subfigure}
\usepackage{float}
\usepackage[sort&compress]{natbib}
\usepackage[dvips]{epsfig}
\usepackage{hhline}
\usepackage{multirow}
\def\ga         {\alpha}

\def\gc         {\gamma}
\def\gC         {\Gamma}
\def\gd         {\delta}
\def\gD         {\Delta}
\def\gee        {\epsilon}
\def\gl         {\lambda}

\def\go         {\omega}
\def\gO         {\Omega}
\def\gr         {\rho}
\def\gs         {\sigma}

\def\gT         {\Theta}
\def\gt         {\theta}
\def\gu         {\tau}
\def\gp         {\phi}

\def\FF		{{\mathbf F}}

\def\RR		{{\mathbf R}}

\def\pp		{{\mathbf p}}

\def\rr		{{\mathbf r}}

\def\KK		{{\mathbf K}}
\def\QQ		{{\mathbf Q}}

\def\uu		{{\mathbf u}}

\def\kk		{{\mathbf k}}
\def\qq		{{\mathbf q}}

\newcommand{\di}{\mathrm{d}}
\newcommand{\im}{\mathrm{i}}
\def\bgc{\mbox{\boldmath $\xi$}}

\def\calgP{\mbox{$\mathit{\Pi}$}}

\def\callD{\mbox{$\mathcal{D}$}}

\def\callG{\mbox{$\mathcal{G}$}}

\def\callL{\mbox{$\mathcal{L}$}}

\def\callQ{\mbox{$\mathcal{Q}$}}

\def\callT{\mbox{$\mathcal{T}$}}

\newcommand{\amend}[1]{#1}%{\textcolor{BrickRed}{#1}}

\def\ai	        {{\em ab--initio}}

\newcommand{\yambo} {{\normalfont\ttfamily yambo}}

\renewcommand{\[}{\left[}
\renewcommand{\]}{\right]}
\renewcommand{\(}{\left(}
\renewcommand{\)}{\right)}
\def\nl         {\right.\\ \left.}

\def\dg         {\dagger}
\def\rar        {\rightarrow}  

\def\Rar        {\Rightarrow}
\def\lrar       {\Longrightarrow}  
  
\def\lrar       {\leftrightarrow}  
  
\newcommand{\olrar}[1]{\overleftrightarrow{#1}}

\newcommand{\eq}[1]{\begin{align}#1\end{align}}
\newcommand{\ml}[1]{\begin{multline}#1\end{multline}}
\newcommand{\eqg}[1]{\begin{gather}#1\end{gather}}
\newcommand{\seq}[1]{\begin{subequations}#1\end{subequations}}

\newcommand{\mll}[2]{\begin{multline}\label{#1}#2\end{multline}}
\newcommand{\eql}[2]{\begin{align}\label{#1}#2\end{align}}
\newcommand{\eqgl}[2]{\seq{\label{#1}\begin{gather}#2\end{gather}}}
\newcommand{\stkout}[1]{\ifmmode\text{\sout{\ensuremath{#1}}}\else\sout{#1}\fi}
\newcommand{\average}[1]{\left\langle #1 \right\rangle}

\newcommand{\bps}[1]{\biggl[ #1 \biggr]}
\newcommand{\bpr}[1]{\biggl( #1 \biggr)}

\newcommand{\lab}[1]{\label{#1}}

\newcommand{\ul}[1]{\underline{#1}}
\newcommand{\oo}[1]{\overline{#1}}
\newcommand{\e}[1]{Eq.~\eqref{#1}}
\newcommand{\es}[2]{Eq.~\eqref{#1}--\eqref{#2}}
\newcommand{\elab}[2]{Eq.(\ref{#1}#2)}
\newcommand{\tab}[1]{Tab.\ref{#1}}
\newcommand{\fig}[1]{Fig.\ref{#1}}
\newcommand{\figlab}[2]{Fig.\ref{#1}#2}
\newcommand{\evalat}[2]{\left.#1\right|_{#2}}
\renewcommand{\sec}[1]{Section\,\ref{#1}}
\newcommand{\app}[1]{Appendix\,\ref{#1}}
\newcommand{\h}[1]{\hat{#1}}
\newcommand{\ti}[1]{\tilde{#1}}
\newcommand{\wh}[1]{\widehat{#1}}

\renewcommand{\t}[1]{\olrar{#1}}

\newcommand{\ocite}[1]{Ref.\cite{#1}}
\def\grad{\mbox{\boldmath $\nabla$}}

\def\Im{{\rm Im}}
\newcommand{\p}{\prime}           % Differential d
\newcommand{\cnrism} {Istituto di Struttura della Materia and Division of Ultrafast Processes in Materials (FLASHit) of the National Research Council, via Salaria Km 29.3, I-00016 Monterotondo Stazione, Italy}
\newcommand{\etsf} {European Theoretical Spectroscopy Facilities (ETSF)}

\begin{document}

\title{Equilibrium and out--of--equilibrium over--screening free phonon self--energy in realistic materials}

\author{Andrea Marini}
\affiliation{\cnrism}
\affiliation{\etsf}

\date{\today}

\begin{abstract}
In model Hamiltonians, like  Fr\"ohlich's, the  electron--phonon interaction is assumed to be screened from the beginning. The same occurs when this interaction
is obtained by using the state--of--the--art density functional perturbation theory as starting point. In this work I formally demonstrate that these approaches
are affected by a severe over--screening error. By using an out--of--equilibrium Many--Body technique I discuss how to merge the many--body approach with
density--functional perturbation theory in order to correct the over--screening error. A symmetric statically screened phonon self--energy is obtained by down--folding
the exact Baym--Kadanoff equations. The statically screened approximation proposed here is shown to have the same long--range spatial limit of the exact
self--energy and to respect the fluctuation--dissipation theorem.  
The doubly screened approximation, commonly used in the literature, is shown, instead, to be over--screened, to violate several Many--Body properties and to have a wrong
spatial long--range decay. The accuracy of the proposed approximation is tested against the exact solution of an extended model Fr\"ohlich Hamiltonian and it is applied to a
paradigmatic material: MgB$_2$. I find that the present treatment enhances the linewidths by $57 \%$ with respect to what has been previously reported for the
anomalous $E_{2g}$ mode. I further discover that the $A_{2u}$ mode is also anomalous (its strong coupling being completely quenched by the over-screened
expression). The present results deeply question methods based on state--of--the--art approaches and impact a wide range of fields such as thermal conductivity,
phononic instabilities and non--equilibrium lattice dynamics.   
\end{abstract}


\maketitle

\section{Introduction}\lab{sec:intro}
%==============================================================================================================================================================
The research on the physics induced by electron--phonon\,(e--p) interaction is one of the most prolific topics in materials science and solid state
physics. A recent review by F.\,Giustino~\cite{Giustino2017} witnesses the countless implications of this interaction on the physics of
electrons and phonons, at and out--of equilibrium. 

The research connected to the e--p interaction that is relevant to the present work can be roughly divided in two main areas: the e--p effect on the
electronic and phononic dynamics. In the first family of applications we find the case of the e--p induced renormalization of the electronic gap
that has been extensively studied from the theoretical~\cite{VanLeeuwen2004,Marini2015} and numerical side~\cite{Ponce2014}. We can conclude that most of the
fundamental aspects are now clear, even if there is still an active research activity about on how to go beyond the state--of--the--art 
approximations and describe more advanced effects, like self--trapping~\cite{LafuenteBartolome2022}.

The case of the e--p effects on the phonon states is very different. The main reason is that, as it will be clear in the following, while the electrons are
natural quantum objects, phonons do not appear in the full Many--Body\,(MB) Hamiltonian written in terms of quantized electrons and nuclei. 
This, indeed, has to be expanded in powers of the atomic displacements that, in turn, define the elemental phonon states. If, however, we look at this procedure
from a  MB point of view we see that the electrons will react to the atomic displacement and the phonon definition will depend on the electronic
response. However, the electronic response emerges dynamically from the solution of the MB problem, while phonons appear directly in the
initial Hamiltonian.

This intrinsic difficulty of defining a coherent approach to the phonon physics is reflected in the different theoretical and methodological approaches that have been proposed in the
literature. We have model Hamiltonians where phonons are introduced as exact bosons from the beginning and the e--p interaction is defined
externally on the basis of physical arguments. A well known example is the Fr\"{o}hlich  Hamiltonian~\cite{Frohlich_1954,Langreth1964,Engelsberg1963}.  In the
case of model Hamiltonians the problem of describing the electronic response to the atomic displacement is ignored by definition. 

Another approach is based on Density Functional Theory\,(DFT) and Density Functional Perturbation Theory\,(DFPT)~\cite{Baroni_1987,Gonze_1995a,Gonze_1995b}.
DFT and DFPT are two electronic density based theories where the description of the electronic screening, connected to the electronic charge oscillation
naturally appears. Indeed DFPT represents the state--of--the art approach to describe phonon frequencies with results in excellent agreement with the
experiments~\cite{Stefano2001}. DFPT is based on the Born--Oppenheimer\,(BO) and adiabatic approximations and, more importantly, within DFT and DFPT the atoms
are treated classically and the Hamiltonian depends parametrically from their positions. The BO approximation allows to decouple the electronic and nuclear
dynamics and define the phonons as oscillations of the atoms around the minimum of the BO energy surface. The adiabatic approximation, instead, assumes that the
electrons follow adiabatically the nuclear oscillations.  An extension of DFPT to include non--adiabatic effects has been proposed in \ocite{Calandra2010}.

The last approach is based on Many--Body Perturbation Theory\,(MBPT)~\cite{Leeuwen2013} where the e--p problem can be formally solved exactly by means of diagrammatic
methods~\cite{Marini2015}. MBPT, however, requires the definition of an initial, reference Hamiltonian and, as I will discuss shortly, this step can potentially
lead to over--screening effects if not properly done. Especially if DFPT is used as starting point.

In this work I present a detailed and complete MB theory of the phonon--self energy 
starting from the fully quantistic and bare electron--nuclei Hamiltonian. By working on the Keldysh contour I propose a derivation that
is valid at the equilibrium and out--of--the equilibrium. I derive a full set of equation of motions for the
$2\times 2$ phonon displacement/momentum  propagators and I show how to derive a symmetric equilibrium form.
By using this I introduce a static--screening approximation and show that it correctly respects the
diagrammatic structure. The doubly--screened approximation for the phonon self--energy, commonly used in the literature, is discussed and compared with the 
present static--screening approximation. The diagrammatic analysis reveals an exploding number of double--counted diagrams and, more importantly,
the exact solution of a generalized model Frh\"{o}lich Hamiltonian clearly show that the doubly--screened 
phonon self--energy \amend{has a wrong vanishing momentum limit that leads to a large underestimation of the long  wave--length phonon widths.}
%behaves as $q^2$ for momentum $q$ going to zero while the exact self-energy remains finite.
I conclude this work by applying the  static--screening approximation to a paradigmatic material: MgB$_2$.

The paper is organized as follows: in the following two sections I introduce an heuristic explanation of the physical origin
of the over--screening error\,(\sec{sec:intro.H}) and a review of the existing literature\,(\sec{sec:intro.review}).

The main body of the work starts in \sec{sec:H} with a careful derivation of the electron--nuclei Hamiltonian by Taylor expanding the fully quantized
\ai\, Hamiltonian. In \sec{sec:ref} I discuss the reference phonon basis and the corresponding electron--nuclei potentials needed to avoid
double--counting terms that appear already in the Hamiltonian. I them move to the MBPT approach that is reviewed in \sec{sec:MBPT}. 
In order to derive  the equation of motion for the atomic displacement and momentum operators (\sec{sec:U}) and for the corresponding Green's function (\sec{sec:D})
I introduce several key concepts like: the vertex function in \sec{sec:vertex} and the phonon self--energy in \sec{sec:pi}. 

In \sec{sec:eq_regime} I mathematically introduce the equilibrium regime deriving a symmetric form of the Dyson equation suitable to introduce 
the static screening approximation. After further simplifications in  \sec{sec:theo_implementation} to obtain a form of the self--energy that will be later
implemented,
I move to the discussion of the key role played by the electronic
screening~(\sec{sec:SCR}). The partially screened and doubly screened approximations are discussed from a diagrammatic point of view in \sec{sec:SS}. In
\sec{sec:jell}, instead, they are compared with the self--energy derived analytically in an exactly solvable model.

In \sec{sec:merge} I discuss how to merge MBPT with methods based on the Born--Oppenheimer approximation, of which DFPT is an example. By using this scheme, in \sec{sec:mgb2}, I calculate the phonon widths of  MgB$_2$.
The paper is concluded, in \app{sec:code} by describing the new code implementation I developed in this work using the Quantum Espresso and Yambo codes, and
providing a scheme of the calculation flow\amend{\,(\app{sec:code.flow})}.

\subsection{The over--screening error: an heuristic introduction}\lab{sec:intro.H}
%--------------------------------------------------------------------------------------------------------------------------------------------------------------
The origin of the over--screening effect can be heuristically introduced by starting from a well--known concept: the Hartree potential.

Let's start from a purely electronic initial Hamiltonian  
\eql{eq:intro.1}{
  \h{H}=\h{H}_e+\h{H}_{e-e},
}
with $\h{H}_{e-e}=\frac{1}{2}\sum\nolimits'_{ij} v\(\h{\rr}_i-\h{\rr}_j\)$ and $v$ the {\bf bare} Coloumb interaction.
The $e-e$ interaction produces, at the lowest order of perturbation theory, a mean--field and classical potential: the Hartree potential
\eql{eq:intro.2}{
  V_H\(\rr\)=\int \di \rr^\p v\(\rr-\rr^\p\) \gr\(\rr^\p\),
}
with $\gr$ the electronic density. In \e{eq:intro.2} the Coulomb interaction is bare. In 1988, Allen, Cohen\&Penn~\cite{PhysRevB.38.2513},
have demonstrated, by using a variational argument, that two test charges in a solid interact through a screened Coulomb interaction, $W=\gee^{-1} v$, with $\gee$ 
the dielectric function of the material.
At the same time it is well known that by  replacing $v$ with $W$ in \e{eq:intro.1} we would get an over--screened Hartree, classical potential. This would be nonphysical and 
makes impossible to rewrite $\h{H}$ in terms of $W$ only. $\h{H}$ must be written in terms of $v$ with its electronic screening being induced by the dynamical
solution of the Many--Body problem.

In the e--p case the situation is the same. Let's add to \e{eq:intro.1}  the e--p interaction:
\eql{eq:intro.3}{
  \h{H}=\h{H}_{ph}+\h{H}_e+\h{H}_{e-p}+\h{H}_{e-e},
}
where $\h{H}_{ph}$ is the non-interacting phonon system taken in the harmonic approximation with eigenvalues $\go_{\nu}$.
$\h{H}_{e-p}=\sum_{ij\nu}g^{\nu}_{ij}\h{\rho}_{ij}\h{u}_{\nu}$ is the electron--phonon interaction with
$g^{\nu}_{ij}$ the {\bf bare} electron--phonon potential, $\h{u}$ the phonon displacement and $\h{\gr}_{ij}$ the electronic density operator.

As I will demonstrate in this work the e--p interaction produces, at the same perturbation order of the Hartree
potential, a mean--field potential $U^\nu\propto g^{\nu}$, \elab{eq:vertex.1}{b}. 
It is also well known\cite{Giustino2017,VanLeeuwen2004,Marini2018} that the variation of the Hartree \amend{potential} screens $g^\nu$. As in the purely
electronic case this means  that any observable evaluated on the interacting ground state of $\h{H}$ will be written in terms of $g^\nu$ {\bf and} $\evalat{g^\nu}{SCR}\sim \gee^{-1} g^\nu$. 
This applies to the electronic gap, optical absorption and phonon energies and related properties.
In practice this means that, in general, {\bf it is not possible} to write
\eql{eq:intro.4}
{
 \h{H}_{e-p}=\evalat{g^{\nu}_{ij}}{SCR} \h{\rho}_{ij}\h{u}_{\nu},
}
when $\h{H}$ contains the e--e interaction, even at a mean--field level. In this case
\e{eq:intro.4} inevitably produces an over--screened error.

\subsection{Literature review}\lab{sec:intro.review}
%--------------------------------------------------------------------------------------------------------------------------------------------------------------
In 1972 P.B.Allen published a work about super--conductivity~\cite{Allen1972} where he derived a relation between the Eliashberg function $\ga^2 F\(\go\)$
and  the phonon widths\,(Eq. 9 of \ocite{Allen1972}). As the $\ga^2 F\(\go\)$ is known to be proportional to $\(\evalat{g^\nu}{SCR}\)^2$
Allen formulation suggested that also the phonon widths have the same proportionality. Allen's expression for the Eliashberg function 
has been applied to evaluate the super--conductive properties of materials in, for example, \ocite{Giustino2007,Errea2015}.

I will explain in \sec{sec:closing.SC} that the Allen's expression is affected by a conceptual error, as it is true only if  \e{eq:intro.4}  is used. This
limitation was already pointed out by Allen himself in 1983\cite{ALLEN19831} where he wrote: {\it The essence of the Fr\"ohlich Hamiltonian is that Coulomb
interactions have already renormalized the electronic energies and e--p interaction, and both Coulomb and electron--phonon interactions have renormalized the
phonon frequencies. 
\amend{
This Hamiltonian is now used to construct the electronic self--energy. It cannot be used to construct the phonon self--energy
because the phonon frequency is already the observed renormalized spectrum.}}

Nevertheless, the expression of the phonon widths proportional to $\(\evalat{g^\nu}{SCR}\)^2$  has been extensively used in the literature to calculate a
wealth of properties. This is especially true in DFT and DFPT based methods where $\evalat{g^\nu}{SCR}$ is a natural by--product of any adiabatic phonon
calculations. 
Nevertheless, within DFT and DFPT the e--e interaction is replaced with the Kohn--Sham\,(KS) Hartree plus exchange--correlation potential. As this potential includes the
Hartree terms it follows that the  e--p interaction is screened.  Indeed within DFPT phonon frequencies are adiabatically renormalized by the electronic
response. Despite this many authors have calculated phonon--related properties by using \e{eq:intro.4} and, thus, over--screening the e--p interaction.
Over--screened calculations have been performed of: (i) the lattice thermal conductivity and transport\cite{Liao_2015,Wang_2016,Tong_2019,Gold-Parker_2018},
(ii) non--adiabatic phonon corrections, line--widths and Kohn
anomalies\cite{Calandra2005,dAstuto2007,Piscanec2004a,Lazzeri2006,Caudal2007,Lazzeri2005,Calandra_2007,Saitta2008,Ferrante2018,Shukla2003,Calandra2010,Novko2018,Nomura2015},
out--of--equilibrium phonon dynamics~\cite{Tong_2020,Caruso_2021,Tanimura2016,Baldini2017,Novko2020}.

While MBPT studies~\cite{Keating1968,VanLeeuwen2004,Marini2018} provide a coherent scheme  in
the DFPT community there is no unique consensus on the procedure to use to correctly screen the e--p interaction. While some 
authors~\cite{Berges2020,Novko2020a,Caruso2017,Campi2021} have proposed a partially screened form of the dynamical matrix, in \ocite{Calandra2010}
the authors presented a variational arguments in favor of the fully screened formulation.  

\section{Definition of the \ai\, electron--phonon Hamiltonian}\lab{sec:H}
%==============================================================================================================================================================
The starting Hamiltonian is a key ingredient of the entire derivation. Indeed a proper definition of the different potentials that appear once the
nuclear and electron--nuclei interaction are Taylor expanded is mandatory to have a well--defined e--p Hamiltonian. In this section I 
review and extend to the phonon case the procedure introduced in  \ocite{Marini2015}.

I start from the generic form of the total Hamiltonian of the system, that I divide in electronic $\h{H}_e$,
nuclear $\h{H}_n$ and electron--nucleus\,(e--n) $\h{H}_{e-n}$ contribution. I keep all components, electrons and nuclei, quantized:
\eq{
\wh{H}=\h{H}_e+\h{H}_n+\h{H}_{e-n}.
\lab{eq:H.1}
}
The electronic and nuclear parts are divided in a kinetic $\widehat{T}$ and interaction part $\wh{W}$:
\seq{
\lab{eq:H.2}
\eqg{
 \h{H}_e =\h{T}_e+\h{H}_{e-e},\\
 \h{H}_n =\h{T}_n+\h{H}_{n-n}.
}}
Note that the nuclear kinetic energy depends on the nuclear momenta.  In the above definitions, the operators are
bare (un--dressed).  

The explicit expression for the bare (e--n) interaction term is
\eq{
\h{H}_{e-n}= -\sum_{I,i}Z_I v\(\h{\rr}_i-\h{\RR}_I\)= \sum_{I,i} V_{e-n}\(\h{\rr}_i,\h{\RR}_I\),
\lab{eq:H.3}
}
where $\h{\RR}_{I}$ is the nuclear position operator for the $I$--th nucleus, $Z_I$ is the corresponding charge, $\h{\rr}_i$ is the electronic position operator of the electron $i$ and
$v\(\rr-\rr'\)=|\rr-\rr'|^{-1}$ is the bare Coulomb potential. Similarly,
\seq{
\lab{eq:H.4}
\ml{
\h{H}_{n-n}=\frac{1}{2}\sum_{I,J}\nolimits' Z_I Z_Jv\(\widehat{\RR}_{I}-\widehat{\RR}_{J}\)=\\
 \frac{1}{2}\sum_{I,J}\nolimits' V_{n-n}\(\h{\RR}_I,\h{\RR}_J\),
}
and,
\eq{
\h{H}_{e-e}=\frac{1}{2}\sum_{i j}\nolimits'v\(\widehat{\rr}_i-\widehat{\rr}_j\),
}}
with $\sum_{ij}\nolimits'=\sum_{i\neq j}$. 

I now split the nuclear position operator $\h{\RR}_I$ in reference and displacement
\eql{eq:H.5}{
 \h{\RR}_I=\overline{\RR}_I+\h{\uu}_I,
}
I now use the notation $\overline{O\(\h{\RR}\)}$, to indicate a quantity or an operator that is evaluated 
with the nuclei frozen in their reference crystallographic positions ($\overline{\RR}$). 

Note that the reference atomic positions are not restricted to correspond to the equilibrium lattice geometry. A more formally correct definition of
equilibrium condition will be given in \sec{sec:ref} when I will define the reference residual atomic force and dynamical matrix.

At this point I can formally introduce the electron--phonon interaction be Taylor expanding \e{eq:H.4}  up to second order in the 
quantized nuclear displacements:
\mll{eq:H.6}
{
 \h{H}_{kind}=\oo{\h{H}}_{kind}+\sum_{I}\oo{\grad_I \h{H}_{kind}} \cdot \h{\uu}_I+\\ \frac{1}{2}\sum_{I,J}\nolimits'\h{\uu}_I\cdot\oo{\grad_I\grad_J \h{H}_{kind}} \cdot
\h{\uu}_J,
}
where $kind=e-n,n-n$. While in the $e-n$ case $\oo{\h{H}_{kind}}$ is an operator, in the $n-n$ case it is a C--number.

In order to define an Hamiltonian suitable to apply MBPT we need a reference basis for the phonon modes. The procedure to introduce this reference
is explained in Sec.III--IV of \ocite{Marini2015}. In the following I review and extend it to the present context. 

I start by introducing a reference tensorial dynamical matrix $\t{C}^{ref}_{IJ}$ which, in turns, define a reference Hamiltonian
\eq{
\lab{eq:H.7}
 \h{H}_{ref}=\frac{1}{2}\sum_{IJ}\nolimits^\p \h{\uu}_I\cdot \t{C}^{ref}_{IJ} \cdot \h{\uu}_J.
}
\e{eq:H.7} defines the corresponding phonon basis via standard canonical transformation. Indeed if $\bgc_{I\gl}$ is the rotation matrix the diagonalize
$\t{C}^{ref}$ we get 
\eql{eq:H.8}
{
 \sum_{JL} \bgc_{J\gl}^T \cdot \frac{\t{C}^{ref}_{JL}}{\sqrt{M_J M_L}} \cdot \bgc_{L\gl^\p}=\gd_{\gl\gl^\p}\go_\gl^2.
}
To keep the notation compact in this section I will use $\gl$ to indicate a phonon branch and momentum. Thanks to \e{eq:H.8} I can define the reference phonon
displacement and momentum operators as components of a vectorial operator $\h{\gp}$:
\eql{eq:H.8.1}
{
\h{\gp}_{s\gl}=
\begin{cases}
  \frac{1}{\sqrt{2}}\(\h{b}^{\dg}_{\gl}+\h{b}_{\gl}\) & s=+\\
  \frac{i}{\sqrt{2}}\(\h{b}^{\dg}_{\gl}-\h{b}_{\gl}\) & s=-
\end{cases}
}
with $\hat{b}^{\dag}_{\gl}$ the phonon $\gl$ creation operator.  It follows that 
\eqgl{eq:H.9}
{
 \h{\uu}_I=\sum_\gl \frac{1}{\sqrt{M_I\go_\gl}} \bgc_{I\gl} \h {\gp}_{+\gl},\\
 \h{\pp}_I=\sum_\gl \sqrt{\frac{\go_\gl}{M_I}} \bgc_{I\gl} \h {\gp}_{-\gl}.
}
Thanks to \e{eq:H.9} it is possible to write that
\amend{
\eq{
\lab{eq:H.10}
\h{T}_{n}+\h{H}_{ref}=\sum_{s\gl}\frac{\go_{\gl}}{2}\hat{\gp}^{\dag}_{s\gl}\hat{\gp}_{s\gl}.
}}

The last step we need to introduce the e--p Hamiltonian is to move in second quantization
\mll{eq:H.10.1}
{
 \h{H}_{e-p}=\h{H}_{e-n}-\oo{\h{H}_{e-n}}=\\\sum_I \int\di\rr_1 \h{\gr}\(\rr_1\) \grad_I
V_{e-n}\(\rr_1,\oo{\RR}_I\)\cdot\h{\uu}_I,
}
where $\h{\gr}\(\rr\)=\h{\psi}^{\dag}\(\rr\)\h{\psi}\(\rr\)$ and $\h{\psi}\(\rr\)=\frac{1}{\sqrt{N}}\sum_{i}\phi_i\(\rr\) \h{c}_i$, with $N$ the number of 
point used to sample the Brillouin zone\footnote{I assume here to consider an extended system}, and $\phi_i\(\rr\)$ the reference electronic wave--function of
the state $i$.

I now use \e{eq:H.9} to write the e--p interaction term in the reference phonon basis
\eql{eq:H.10.2}
{
 \h{H}_{e-p}-\oo{\h{H}_{e-n}}=\sum_\gl \int\di\rr_1 \h{\gr}\(\rr_1\) g^\gl\(\rr_1\) \h{\gp}_{+\gl},
}
with
\eql{eq:H.10.3}
{
 g^\gl\(\rr_1\)= \sum_I \frac{1}{\sqrt{M_I\go_\gl}} \grad_I V_{e-n}\(\rr_1,\oo{\RR}_I\)\cdot \bgc_{I\gl}
}

Thanks to Eqs.(\ref{eq:H.10.1}--\ref{eq:H.10.3}) we can finally Taylor expand $\h{H}$:
\begin{widetext}
\eq{
\lab{eq:H.11}
\amend{\wh{H}}=  \sum_{i} \gee_{i} \hat{c}^{\dagger}_{i} \hat{c}_{i}+ \wh{H}_{e-e}+
\sum_{\gl}\bps{\frac{\go_{\gl}}{2}\sum_s\(\h{\gp}_{s\gl}^\dag\h{\gp}_{s\gl}\)+\bpr{\Xi_{\gl}+\sum_{ij} g_{ij}^{\gl} \h{\gr}_{ij}} \h{\gp}_{+\gl}}+
\sum_{\gl\gl^\p}\bpr{ \sum_{ij} \gt_{ij}^{\gl\gl^\p}\h{\gr}_{ij}+\gT_{\gl\gl^\p}} \h{\gp}_{+\gl} \h{\gp}_{+\gl^\p}-\h{H}^{ref}.
}
\end{widetext}
\e{eq:H.11} is the complete form of the quantized electron--phonon Hamiltonian. I have introduced the electronic density matrix operator: $\h{\gr}_{ij}=
\hat{c}^{\dagger}_{i} \hat{c}_{j}$.
The different interaction potentials introduced in \e{eq:H.11} are 
\seq{
\lab{eq:H.12}
\eqg{
g^{\gl}_{ij}=\bra{i} g^{\gl}\(\rr\) \ket{j},\\
\Xi_{\gl}= \frac{1}{2}\sum_{I,J}\nolimits' \partial_{\gl} V_{n-n}\(\oo{\RR}_I,\oo{\RR}_J\),\\
\gt_{nm\kk}^{\gl\gl^\p}= \frac{1}{2} \sum_I \bra{i} \partial^2_{\gl\gl^\p} V_{e-n}\(\rr,\oo{\RR}_I\) \ket{j},
}
and
\eq{
\gT_{\gl\gl^\p}= \frac{1}{2}\bps{\sum_{I,J}\nolimits'\partial^2_{\gl\gl^\p} V_{n-n}\(\oo{\RR}_I,\oo{\RR}_J\) -\gd_{\gl\gl^\p}\go_\gl}.
}
}
In \e{eq:H.12} the definition of the derivative along the phonon $\gl$ easily follows from \elab{eq:H.9}{a}.

In the case of $\gt_{ij}^{\gl\gl^\p}$ it is convenient to introduce the corresponding real--space function
\eql{eq:H.14}
{
\gt_{\gl\gl^\p}\(\rr\)= \frac{1}{2} \sum_I   \partial^2_{\gl\gl^\p} V_{e-n}\(\rr,\oo{\RR}_I\).
}

\subsection{The reference atomic force and self--energy}\lab{sec:ref}
%==============================================================================================================================================================
I can now inspect the physical meaning of the different potentials appearing in \e{eq:H.11} in order to arrive to a more compact form. Indeed we start by observing
that
\mll{eq:ref.1}
{
\Xi_{\gl}+\sum_{ij}g_{ij}^\gl\average{\h\gr_{ij}}=\frac{1}{2}\sum_{I,J} \oo{\partial_{\gl} V_{n-n}\(\RR_I,\RR_J\)}+\\
\sum_I\int\di\rr \oo{\partial_\gl V_{e-n}\(\rr,\RR_I\)} \gr\(\rr\)=-F_\gl.
}
$F_\gl$ is the force acting on the atoms along the $\gl$ phonon direction. \e{eq:ref.1} makes clear that $F_\gl$ is a functional of the electronic density,
$\gr\(\rr\)$.  \e{eq:ref.1} mathematically defines the equilibrium condition for the \ai\, Hamiltonian, \e{eq:H.11}: 
the atomic positions $\oo{\RR}_I$ correspond to an equilibrium configuration if the corresponding electronic density is such that $F^{ref}_\gl\[\gr\]=0$.
We will see in the \sec{sec:merge} the implications of using, as reference, the results of a DFPT calculation.

The second term is
\mll{eq:ref.2}
{
\sum_{ij}\gt_{ij}^{\gl\gl^\p}\average{\h{\gr}_{ij}}+\gT_{\gl\gl^\p}-\gd_{\gl\gl^\p}\frac{\go_\gl}{2}=\\
\frac{1}{2}\bps{\sum_I\int\di\rr \oo{\partial^2_{\gl\gl^\p} V_{e-n}\(\rr,\RR_I\)} \gr\(\rr\)+ \\+
\sum_{I,J}\oo{\partial^2_{\gl\gl^\p} V_{n-n}\(\RR_I,\RR_J\)} -\gd_{\gl\gl^\p}\go_\gl}=
-\frac{1}{2}C^{ref}_{\gl\gl^\p}.
}
\e{eq:ref.2} defines $C^{ref}$ that corresponds to the e--n contribution to the dynamical matrix when the atoms sit in
their reference configuration.

Thanks to \e{eq:ref.1} and \e{eq:ref.2} we obtain that
\mll{eq:ref.3}{
\amend{\wh{H}}=  \sum_{i} \gee_{i} \hat{c}^{\dagger}_{i} \hat{c}_{i}+ \wh{H}_{e-e}+\sum_{\gl}\bps{\frac{\go_{\gl}}{2}\sum_s
\(\h{\gp}_{s\gl}^\dag\h{\gp}_{s\gl}\)+\\+\(\h{\callL}_\gl+ \sum_\mu \h{\callQ}_{\mu\gl} \h{\gp}_{+\mu}\) \h{\gp}_{+\gl} }
}
where
\eqgl{eq:ref.4}
{
 \h{\callL}_\gl= \sum_{ij} g_{ij}^{\gl} \Delta\h{\gr}_{ij}-F_\gl,\\
 \h{\callQ}_{\gl\mu}=\sum_{ij} \gt_{ij}^{\gl\mu} \Delta\h{\gr}_{ij}-\frac{1}{2} C^{ref}_{\gl\mu},
}
and $\Delta\h{\gr}_{ij}=\h{\gr}_{ij}-\average{ \h{\gr}_{ij} }$.

Let's keep in mind that \e{eq:ref.4} can be equivalently expressed in terms of $\h{\gr}\(\rr\)$
\eqgl{eq:ref.5}
{
 \h{\callL}_\gl= \int \di \rr g^{\gl}\(\rr\) \Delta \h{\gr}\(\rr\)-F_\gl,\\
 \h{\callQ}_{\gl\mu}=\int \di \rr \gt^{\gl\mu}\(\rr\) \Delta\h{\gr}\(\rr\)-\frac{1}{2} C^{ref}_{\gl\mu}.
}

\section{Review of the Many--Body Perturbation Theory approach to the Phonon problem}\label{sec:MBPT}
%==============================================================================================================================================================
In this Section I review the basic steps in the derivation of the MBPT phonon self--energy by using as a starting point the bare and undressed Hamiltonian,
\e{eq:H.11}. 
I will extend the derivation presented in Marini \& Pavlyukh\cite{Marini2018} by introducing the reference potentials defined in \sec{sec:ref}
and, also, by deriving the usual second--order differential equation for the displacement--displacement Green's function from the first order $2\times 2$ equation of
motion.
Everything is derived by using the Keldysh formalism~\cite{Leeuwen2013} and the functional derivatives method\cite{Strinati1988,VanLeeuwen2004} which is an
approach alternative to the standard diagrammatic method.

\subsection{Equation of motion for the phonon displacement and momentum fields}\lab{sec:U}
%--------------------------------------------------------------------------------------------------------------------------------------------------------------
In order to define the phonon propagator and  derive its equation of motion I need as input the
equations of motions for the elemental fields $\h{\gp}_{s\gl}$. 
First I switch to the time-dependent Heisenberg operatorial representation with time arguments $z$ on the Keldysh contour\cite{Leeuwen2013}. 
\amend{The Keldysh formalism ensures that the present results are not limited to the equilibrium and/or zero-temperature cases}.

I now apply Heisenberg's EOM for operators, 
\eq{
\lab{eq:U.1}
\amend{\frac{\di}{\di z}\h{O}(z)=i \bps{\h{O}\(z\),\h{H}\(z\)}_{-}}.
}
By using the fact that
\eqgl{eq:U.2}
{
 \[\h{\gp}_{s\gl},\h{\gp}_{s^\p\mu}\]_{-}=\[\underline{\gs}_2\]_{ss^\p} s \gd_{\gl\mu},\\
 \[\h{c}_i,\h{c}_j\]_{+}=\h{c}_i\h{c}_j+\h{c}_j\h{c}_i=\gd_{ij},
}
it follows that 
\eqgl{eq:U.3}
{
\frac{\di}{\di z}\h{\gp}_{-\gl}\(z\)=  -\go_\gl\h{\gp}_{+\gl}\(z\)-\h{\callL}_\gl\(z\)-2\sum_\mu\h{\callQ}_{\gl\mu}\h{\gp}_{+\mu}\(z\),\\
\frac{\di}{\di z}\h{\gp}_{+\gl}\(z\)=  \go_\gl\h{\gp}_{-\gl}\(z\).
}
In \e{eq:U.2} $\underline{\gs}_2$ is the  $2\times 2$ Pauli matrix
\eql{eq:U.4}
{
\underline{\gs}_2=
 \begin{pmatrix}
   0 & -\im\\
   \im & 0
 \end{pmatrix}.
}

\subsection{The phonon propagator matrix}\lab{sec:D}
%--------------------------------------------------------------------------------------------------------------------------------------------------------------
Having then obtained the coupled EOMs for the $\h{\phi}_{s\gl}$ fields, we combine them to arrive at the EOM for the phonon Green's function\,(GF) matrix $D^{s_1s_2}_{\gl_1\gl_2}\(z_1,z_2\)$:
\eq{
 \lab{eq:D.1}
  D^{s_1s_2}_{\gl_1\gl_2}\(z_1,z_2\) = \(-\im\) \average{\mathcal{T}\{\gD\h{\gp}_{s_1\gl_1}\(z_1\) \gD\h{\gp}^\dag_{s_2\gl_2}\(z_2\) \}},
}
where $\expval{\cdots}$ is the trace over the exact density matrix, $\mathcal{T}$ the contour-ordering operator and
$\gD\h{\gp}_{s\gl}=\h{\gp}_{s\gl}-\average{\h{\gp}_{s\gl}}$.
The electronic GF is
\eq{
 \lab{eq:D.2}
  G\(1,2\)=\(-\im\) \expval{\mathcal{T}\{ \h{\psi}\(1\) \h{\psi}^\dg\(2\) \}},
}
with $1=\(\rr_1,z_1\)$.

The phonon  GF (and self--energy) can be represented as $2\times 2$ matrices, $\ul{D}_{\gl_1\gl_2}\(z_1,z_2\)$. 
Its EOM  can be obtained by introducing a fictitious time--dependence in $H$, as described in \ocite{Marini2018}:
\eql{eq:D.3}
{
 \amend{\h{H}_{\xi\eta}\(z\)=\h{H}+\sum_{s\gl} \xi_{s\gl}\(z\)\h{\gp}_{s\gl}+\int \di \rr_1 \eta\(1\)\h{\gr}\(\rr_1\).}
}
Thanks to the introduction of $\xi_{s\gl}\(z\)$ it is straightforward to demonstrate that
\eql{eq:D.4}
{
 D^{s_1s_2}_{\gl_1\gl_2}\(z_1,z_2\)=\frac{\gd \average{\h{\gp}_{s_1\gl_1}\(z_1\)}}{\gd \xi_{s_2\gl_2}\(z_2\)}.
}
\amend{In \e{eq:D.4} the average is evaluated using  $\h{H}_{\xi\eta}\(z\)$.  In the following all averages are dependent on $\xi$ and}
$\eta$ and the limit $\eta,\xi\rar 0$ will be performed at the end of the derivation.

From \e{eq:D.4} and \e{eq:U.3} it follows that
\begin{widetext}
\seq{
\lab{eq:D.5}
\eq
{
 \im\frac{\di}{\di z_1}
D^{s_1s_2}_{{\gl_1}{\gl_2}}\(z_1,z_2\)=\bps{\ul{\gs}_2\bpr{\go_{\gl_1}\ul{D}_{{\gl_1}{\gl_2}}\(z_1,z_2\)+\gd_{z_1z_2}\gd_{\gl_1\gl_2}}}_{s_1s_2}+
\gd_{s_1 -}\frac{\gd \average{\h{\callL}_{\gl_1}\(z_1\)+2\sum_{\gl_3}\h{\callQ}_{{\gl_1}\gl_3}\(z_1\)\h{\gp}_{+\gl_3}\(z_1\)}}{\gd \xi_{s_2\gl_2}\(z_2\)}.
}
Similarly we can evaluate the right derivative
\eq
{
 \im\frac{\di}{\di z_2}
D^{s_1s_2}_{{\gl_1}{\gl_2}}\(z_1,z_2\)=\bps{\bpr{\go_{\gl_2}\ul{D}_{{\gl_1}{\gl_2}}\(z_1,z_2\)+\gd_{z_1z_2}\gd_{\gl_1\gl_2}}\ul{\gs}_2}_{s_1s_2}+
\gd_{s_2 -}\frac{\gd \average{\h{\callL}^{\dag}_{\gl_2}\(z_2\)+2\sum_{\gl_3}\h{\callQ}^{\dag}_{{\gl_2}\gl_3}\(z_2\)\h{\gp}_{+\gl_3}\(z_2\)}}{\gd \xi_{s_1\gl_1}\(z_1\)}.
}
}
\end{widetext}
In \e{eq:D.5} $\gd_{z_1z_2}=\gd\(z_1-z_2\)$.
We need now to evaluate the r.h.s. of \e{eq:D.5}. Let's consider \elab{eq:D.5}{a} as the right derivative can be similarly worked out. 
There are two terms:
\eql{eq:D.6}
{
 \frac{\gd \average{\h{\callL}_{\gl_1}\(z_1\)}}{\gd \xi_{s_2\gl_2}\(z_2\)}= 
 \int \di \rr_1 g^{\gl_1}\(\rr_1\) \frac{\gd\average{\h{\gr}\(1\)}}{\gd \xi_{s_2\gl_2}\(z_2\)},
}
and 
\mll{eq:D.7}
{
 \frac{\gd \average{\h{\callQ}_{\gl_1\gl_3}\(z_1\)\h{\gp}_{+\gl_3}\(z_1\)}}{\gd \xi_{s_2\gl_2}\(z_2\)}=
 -\frac{1}{2} C^{ref}_{\gl_1\gl_3}D_{\gl_3\gl_2}\(z_1,z_2\)+\\\int \di \rr_1
  \gt^{\gl_1\gl_3}\(\rr_1\) \frac{\gd\average{ \Delta\h{\gr}\(1\) \h{\gp}_{+\gl_3}\(z_1\) }}{\gd \xi_{s_2\gl_2}\(z_2\)}.
}
In evaluating \e{eq:D.6} I have used \elab{eq:ref.4}{a} and the fact that $\frac{\gd \Xi}{\gd \xi}=0$.
Moreover the last term on the r.h.s. of \e{eq:D.7} is, at least, proportional to $\h{u}_{\gl}^2$ via $\gt$. This implies 
that,  within the harmonic approximation
\eql{eq:D.8}
{
 \frac{\gd \average{\h{\callQ}_{\gl_1\gl_3}\(z_1\)\h{\gp}_{+\gl_3}\(z_1\)}}{\gd \xi_{s_2\gl_2}\(z_2\)} \approx 
 -\frac{1}{2} C^{ref}_{\gl_1\gl_3}D^{+s_2}_{\gl_3\gl_2}\(z_1,z_2\).
}

\subsection{The vertex function}\lab{sec:vertex}
%--------------------------------------------------------------------------------------------------------------------------------------------------------------
In order to write \e{eq:D.6} in terms of Green's functions I use the chain rule 
\mll{eq:vertex.0}
{
 \frac{\gd \average{\h{\gr}\(1\)}}{\gd \xi_{s_2\gl_2}\(z_2\)}=-\int\di 34 G\(1,3\)\frac{\gd G^{-1}\(3,4\)}{\gd \xi_{s_2\gl_2}\(z_2\)}G\(4,1\).
}
The Dyson equation for the electronic Green's function is
\mll{eq:vertex.0.1}
{
 G\(1,2\)=G^0\(1,2\)+\int \di 34 G^0\(1,3\)\\\bps{V_{tot}\(3\)\gd_{34}+M\(3,4\)}G\(4,2\),
}
with $V_{tot}\(1\)=V_H\(1\)+\eta\(1\)+U\(1\)$ and 
\eql{eq:vertex.0.2}{
 \gd_{12}=\gd_{z_1z_2}\gd_{\rr_1\rr_2}=\gd\(z_1-z_2\)\gd\(\rr_1-\rr_2\).
} 
In \e{eq:vertex.0.1}
\eqgl{eq:vertex.1}
{
 V_H\(1\)=\int \di \rr_2 v\(\rr_1-\rr_2\)\gr\(2\),\\
 U\(1\)=\sum_{\gl_1\gl_2} g^{\gl_1}\(\rr_1\) \int \di z_2 D^{+s_2}_{\gl_1\gl_2}\(z_1,z_2\) \xi_{s_2\gl_2}\(z_2\).
}
$V_H$ and $U$ are the mean--field electronic potentials induced by the e--e\,(Hartree) and
e--p\,(tad--pole) interactions.

\begin{figure}
 {\centering
 \includegraphics[width=\columnwidth]{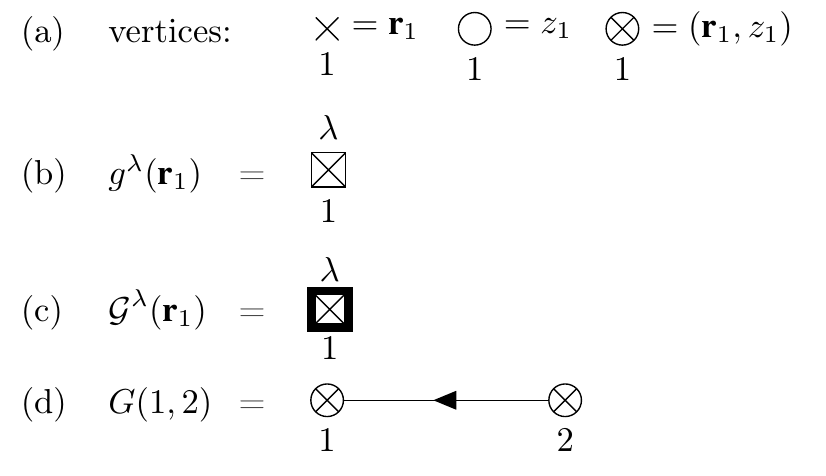}
 }
 \caption{Definition of the diagrammatic elements used in this work. (a) $\bigcirc$ and
 $\times$ represent a generic time and position point respectively.  These two
 symbols can be combined to indicate a time and position vertex $\oplus_1$
 equivalent to $1=\(\rr_1,z_1\)$. (b) A box around a spatial point represents
 the e--p bare potential $g^{\gl}\(\rr_1\)$.
 (c) Dressed e--p potential. (d) Electronic Green's function.\label{fig:1}}
\end{figure}

The mass operator appearing in \e{eq:vertex.0.1} will not be discussed here. I assume it to be the exact 
$M\(1,2\)$. More information can be found, for example, in \ocite{VanLeeuwen2004,Marini2018}.

I can now work out the functional derivative $\frac{\gd}{\gd \xi}$
\eql{eq:vertex.2}
{
 \frac{\gd}{\gd \xi_{s_2\gl_2}\(z_2\)}=\int \di 34 \frac{\gd U\(3\)}{\xi_{s_2\gl_2}\(z_2\)}
\frac{\gd V_{tot}\(4\)}{\gd U\(3\)}
\frac{\gd}{\gd V_{tot}\(4\)}.
}
I now observe that from \elab{eq:vertex.1}{b} it follows
\eql{eq:vertex.3}
{
 \frac{\gd U\(3\)}{\gd \xi_{s_2\gl_2}\(z_2\)}=\sum_{\gl_3} g^{\gl_3}\(\rr_3\) D^{+s_2}_{\gl_3\gl_2}\(z_3,z_2\),
}
while
\mll{eq:vertex.4}
{
 \frac{\gd V_{tot}\(4\)}{ \gd U\(3\)}=\gd_{43}+\int \di 5 v\(4,5\) \frac{\gd \gr\(5\)}{\gd U\(3\)}=\\
 \gd_{43}+\int \di  5 v\(4,5\) \chi\(5,3\)=\gee^{-1}\(4,3\).
}
Thanks to \es{eq:vertex.1}{eq:vertex.4} I can finally introduce the  reducible  and irreducible  e--p vertex functions
\eqgl{eq:vertex.5}
{
 \gC^{s_3\gl_3}\(12,z_3\)\equiv -\frac{\gd G^{-1}\(1,2\)}{\gd \xi_{s_3\gl_3}\(z_3\)},\\
 \ti{\gC}\(12,3\)\equiv -\frac{\gd G^{-1}\(1,2\)}{\gd V_{tot}\(3\)}.
}
The two vertex functions are connected via \e{eq:vertex.2}
\mll{eq:vertex.6}
{
 \gC^{s_3\gl_3}\(12,z_3\)=
 \sum_{\gl_4}
 \int \di 45 
 \ti{\gC}\(12,5\)\\
 \gee^{-1}\(5,4\) 
 g^{\gl_4}\(\rr_4\) D^{+s_3}_{\gl_4\gl_3}\(z_4,z_3\).
}
\e{eq:vertex.6} allows me to introduce the non--local, and time--dependent effective e--p interaction potential
\eql{eq:vertex.7}
{
 \callG^{\gl}\(1,z_2\)=\int \di \rr_2 \gee^{-1}\(1,2\) g^{\gl_1}\(\rr_2\).
}

Before connecting the vertex function to the phonon self--energy we need to derive its equation of motion. This has been already done in \ocite{Marini2018} for the
reducible $\gC$. A similar procedure, based on the chain rule, can be followed for the irreducible $\ti{\gC}$
\mll{eq:vertex.8}
{
 \ti{\gC}\(12,3\)=\gd_{12}\gd_{13}+\int \di 4567 
 \frac{\gd_t M\(1,2\)}{\gd_t G\(4,5\)}\\
 G\(4,6\) G\(7,5\) \ti{\gC}\(67,3\).
}

To conclude this section we have to put together the results of \e{eq:vertex.0} and \es{eq:vertex.3}{eq:vertex.6} to get
\seq{
 \lab{eq:vertex.9}
\ml
{
 \frac{\gd \average{\h{\callL}_{\gl_1}\(z_1\)}}{\gd \xi_{s_2\gl_2}\(z_2\)}= \\ 
 \sum_{\gl_4} \int \di\rr_1 \di 3 \di z_4 \[g^{\gl_1}\(\rr_1\) \ti{\chi}\(1,3\)\callG^{\gl_4}\(3,z_4\)\]\\ D^{+s_2}_{\gl_4\gl_2}\(z_4,z_2\)
}
with
\eq{
 \ti{\chi}\(1,2\)=-\im \int \di 34  G\(1,3\)G\(4,1\) \ti{\gC}\(34,2\).
}}

\subsection{The equation of motion for $\ul{D}$ and the left and right self--energies}\lab{sec:pi}
%--------------------------------------------------------------------------------------------------------------------------------------------------------------
If we now use \e{eq:vertex.9}, \e{eq:D.8} in \e{eq:D.5} we finally arrive at the equation of motion for for $\ul{D}$:
\begin{widetext}
\seq{
\lab{eq:pi.1}
\eq
{
\im\frac{\di}{\di z_1} \ul{D}_{{\gl_1}{\gl_2}}\(z_1,z_2\)=
\ul{\gs}_2\bpr{\go_{\gl_1}\ul{D}_{{\gl_1}{\gl_2}}\(z_1,z_2\)+\gd_{z_1z_2}\gd_{\gl_1\gl_2}}+\evalat{\ul{I}_{\gl_1\gl_2}\(z_1,z_2\)}{L},
}
and
\eq{
-\im\frac{\di}{\di z_2} \ul{D}_{{\gl_1}{\gl_2}}\(z_1,z_2\)=
\bpr{\go_{\gl_2}\ul{D}_{{\gl_1}{\gl_2}}\(z_1,z_2\)+\gd_{z_1z_2}\gd_{\gl_1\gl_2}}\ul{\gs}_2+
\evalat{\ul{I}_{\gl_1\gl_2}\(z_1,z_2\)}{R}.
}
}
where
\seq{
\lab{eq:pi.1.a}
\eq
{
 \evalat{\ul{I}_{\gl_1\gl_2}\(z_1,z_2\)}{L}=
 -\im\int \di z_3 \sum_{\gl_3} \(\Pi^L_{\gl_1\gl_3}\(z_1,z_3\)-\Pi^{ref}_{\gl_1\gl_3}\gd_{z_1z_3}\)
 \begin{pmatrix}
   D^{+-}_{\gl_3\gl_2}\(z_3,z_2\) & D^{++}_{\gl_3\gl_2}\(z_3,z_2\) \\
   0 & 0
 \end{pmatrix}
}
and
\eq
{
 \evalat{\ul{I}_{\gl_1\gl_2}\(z_1,z_2\)}{R}=
 \im\int \di z_3 
 \sum_{\gl_3} 
 \begin{pmatrix}
   D^{-+}_{\gl_1\gl_3}\(z_1,z_3\) & 0 \\
   D^{++}_{\gl_1\gl_3}\(z_1,z_3\) & 0
 \end{pmatrix}
 \(\Pi^R_{\gl_3\gl_2}\(z_3,z_2\)-\Pi^{ref}_{\gl_3\gl_2}\gd_{z_3z_2}\).
}
}
The key ingredients in \es{eq:pi.1}{eq:pi.1.a} are the phonon self--energies, defined as
\seq{
\lab{eq:pi.2}
\eq{
 \Pi^{L}_{\gl_1\gl_2}\(z_1,z_2\)= \int \di\rr_1 \di 3 g^{\gl_1}\(\rr_1\) \ti{\chi}\(1,3\)\callG^{\gl_2}\(3,z_2\),
}
and
\eq{
 \Pi^{R}_{\gl_1\gl_2}\(z_1,z_2\)= \int \di\rr_2 \di 3 \callG^{\gl_1}\(z_1,3\) \ti{\chi}\(3,2\)g^{\gl_2}\(\rr_2\).
}
}
In the following I will use the short notation $\gD \Pi^{L/R}_{\gl_1\gl_2}\(z_1,z_2\)= \Pi^{L/R}_{\gl_1\gl_2}\(z_1,z_2\)-C^{ref}_{\gl_1\gl_2}\gd_{z_1z_2}$.
\end{widetext}

\es{eq:pi.1}{eq:pi.2} represent a crucial result of this work. It demonstrates that, if the MBPT derivation is done starting from an \ai\, Hamiltonian 
the reference atomic positions that define the zero--th order of the harmonic expansion define a term, $\ul{C}^{ref}$ that needs to be
removed from the full MBPT self--energy in order to avoid double--counting of correlation effects. 
This represents the analogous of the 
electronic case where it is well established that the DFT exchange--correlation potential, $V_{xc}$ must be removed from the MBPT electronic
self--energy~\cite{Onida2002}. $\Pi^{ref}$ plays exactly the same role of $V_{xc}$.

\begin{figure}
  {\centering
  \includegraphics[width=\columnwidth]{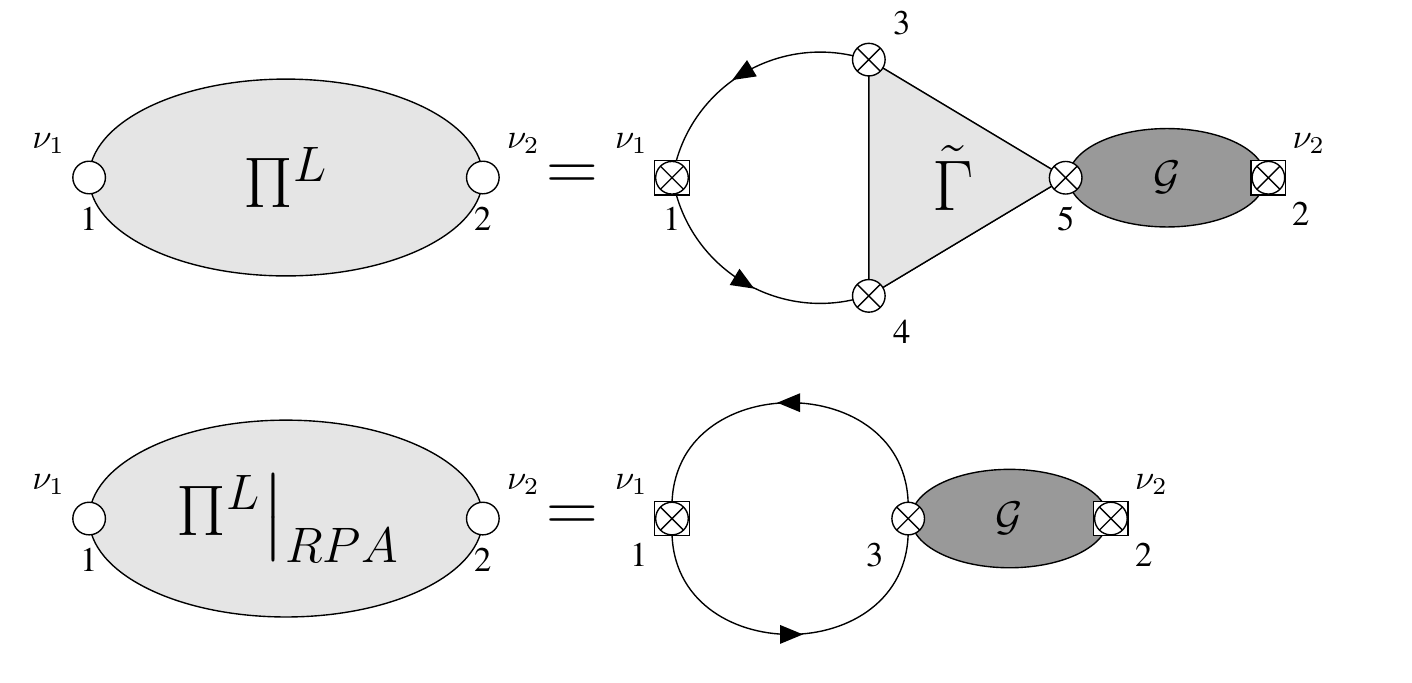}
  }
  \caption{
Diagrammatic representation of $\Pi^L_{\gl_1\gl_2}\(z_1,z_2\)$. Full self--energy\,(upper frame) and
within the RPA approximation corresponding to the approximation $\ti{\gC}=1$\,(lower frame).
\label{fig:2}}
\end{figure}

Let me conclude this section by introducing the diagrammatic representation of \elab{eq:pi.2}{a}. In \fig{fig:1} all ingredients of the diagrammatic
representation are showed and in the upper frame of \fig{fig:2} $\Pi^L_{\gl_1\gl_2}\(z_1,z_2\)$ is diagrammatically represented.  In the lower frame of
\fig{fig:2}, instead, I show the self--energy in the Random--Phase Approximation\,(RPA) that I will discuss in detail in the next sections.

\section{The equilibrium regime}\label{sec:eq_regime}
%==============================================================================================================================================================
\es{eq:pi.1}{eq:pi.2}  are written on the Keldysh contour and, in addition, are first--order time--derivatives. In this section
I will demonstrate how to move in the equilibrium regime defining the corresponding self--energy for the displacement--displacement
component of the phonon Green's function matrix.

As a first step we remind that the real--time components of the Green's function are obtained by applying the Langreth rules\cite{Leeuwen2013}. Here
I am interested in the retarded component of $\ul{D}$ that I will refer to as   $\ul{\callD}\(t_1,t_2\)$.
$\callD$ will in general depend on $t_1$ and $t_2$ and not on $t_1-t_2$. In order to introduce the equilibrium regime we move from 
\eql{eq:reg.1}
{ 
\(t_1,t_2\)\Rar \(T=\frac{\(t_1+t_2\)}{2},\gu=\frac{\(t_1-t_2\)}{2}\).
}
In the $\(T,\gu\)$ basis the equilibrium regime is defined by the condition $\frac{\di}{\di T} \evalat{\ul{\callD}_{\gl_1\gl_2}\(T,\gu\)}{eq}=0$. It follows that
\eql{eq:reg.2}
{ 
\evalat{\ul{D}_{\gl_1\gl_2}\(t_1,t_2\)}{eq}=\ul{D}_{\gl_1\gl_2}\(t_1-t_2\).
}
The goal now is to derive from \es{eq:pi.1}{eq:pi.2} the equation of motion for $\ul{\callD}_{\gl_1\gl_2}\(\gu\)$. In the following I will demonstrate that 
the EOM for $\ul{D}_{\gl_1\gl_2}$  can
be closed in three equivalent formulations in the subspace of $D^{++}_{\gl_1\gl_2}\(\gu\)$ only. Only one of these 
will lead to a symmetrized form suitable to take the static screening limit.

Let's start by the $\(s_1,s_2\)$ components of \es{eq:pi.1}{eq:pi.1.a}. The components of the left and right time derivative are
derived in \app{sec:D_matrix_eom}.

I now define a symmetric differential operator:
\eql{eq:reg.5}
{
 \frac{\di}{\di\gu}=\frac{1}{2}\(\frac{\di}{\di t_1}-\frac{\di}{\di t_2}\).
}
From \es{eq:reg.3}{eq:reg.4} it follows that
\eql{eq:reg.6}
{
\frac{\di}{\di\gu} \callD^{++}_{\gl_1\gl_2}\(\gu\)=\frac{\go_{\gl_1}}{2}  \callD^{-+}_{\gl_1\gl_2}\(\gu\)-\frac{\go_{\gl_2}}{2}
\callD^{+-}_{\gl_1\gl_2}\(\gu\).
}
\begin{widetext}
From Eqs.(\ref{eq:reg.3}b--\ref{eq:reg.3}c) and  Eqs.(\ref{eq:reg.4}b--\ref{eq:reg.4}c) we see that if we apply $\frac{\di}{\di\gu}$ again to \e{eq:reg.6} we
can rewrite the r.h.s. in terms of $\callD^{++}$ and $\callD^{--}$. Indeed
\mll{eq:reg.7}
{
\frac{\di^2}{\di\gu^2}
\callD^{++}_{\gl_1\gl_2}\(\gu\)=-\callD^{++}_{\gl_1\gl_2}\(\gu\)\frac{\(\go_{\gl_1}^2+\go_{\gl_2}^2\)}{4}-\callD^{--}_{\gl_1\gl_2}\(\gu\)\frac{\go_{\gl_1}\go_{\gl_2}}{2}+\\
-\gd\(\gu\)\frac{\(\go_{\gl_1}+\go_{\gl_2}\)}{2}-\frac{1}{2}\int \di \gu^\p \sum_{\gl_3} 
\bps{ 
\go_{\gl_1}
\gD\calgP^L_{\gl_1\gl_3}\(\gu-\gu^\p\)
\callD^{++}_{\gl_3 \gl_2}\(\gu^\p\)+
\go_{\gl_2}
\callD^{++}_{\gl_1 \gl_3}\(\gu-\gu^\p\)
\gD\calgP^R_{\gl_3\gl_2}\(\gu^\p\)
}.
}
\end{widetext}
In \e{eq:reg.7} the last term on the r.h.s. acquires a $2$ prefactor due to the transformation $\di t_3\rar \di \gu^\p$.
The last step we need to close \e{eq:reg.7} is to observe that if we assume that
$\average{\callT\{\h{b}_{\gl_1}\(z_1\)\h{b}_{\gl_2}\(z_2\)\}}=\average{\callT\{\h{b}^\dag_{\gl_1}\(z_1\)\h{b}^\dag_{\gl_2}\(z_2\)\}}=0$ it follows from
\e{eq:D.1} that
\eql{eq:reg.8}
{
 \callD^{--}_{\gl_1 \gl_2}\(\gu\)=\callD^{++}_{\gl_1 \gl_2}\(\gu\).
}
If we use \e{eq:reg.8} assuming also the 
$\callD^{++}_{\gl_1 \gl_2}\(\gu\)\sim \gd_{\gl_1\gl_2}\callD^{++}_{\gl_1}\(\gu\)$ and
$\gD\calgP^{++}_{\gl_1 \gl_2}\(\gu\)\sim \gd_{\gl_1\gl_2}\gD\calgP^{++}_{\gl_1}\(\gu\)$ 
we finally rewrite 
\e{eq:reg.7} as
\begin{widetext}
\mll{eq:reg.9}
{
\frac{\di^2}{\di\gu^2}
\callD^{++}_{\gl_1}\(\gu\)=
-\go_{\gl_1}\gd\(\gu\)
-\go_{\gl_1}^2\callD^{++}_{\gl_1}\(\gu\)
-\frac{\go_{\gl_1}}{2}
\int \di \gu^\p 
\bps{ 
\gD\calgP^L_{\gl_1}\(\gu-\gu^\p\)
\callD^{++}_{\gl_1}\(\gu^\p\)+
\callD^{++}_{\gl_1}\(\gu-\gu^\p\)
\gD\calgP^R_{\gl_2}\(\gu^\p\)
}.
}
\end{widetext}
\e{eq:reg.9} is another crucial result of this work. It defines a symmetric second order equation of motion for the retarded phonon Green's function at the equilibrium where
both the left and right self--energies appear. As it will be clear in the following \e{eq:reg.9} admits a well--defined and formally correct procedure to
introduce the static--screening approximation.

Indeed there are other two forms of the Dyson equation for $D^{++}$ that can be obtained by applying $\frac{\di}{\di t_1}$ to \elab{eq:reg.3}{a} and
$\frac{\di}{\di t_2}$ to \elab{eq:reg.4}{a}:
\seq{
\lab{eq:reg.10}
\ml{
\frac{\di^2}{\di t_1^2}\callD^{++}_{\gl_1}\(t_1-t_2\)=
-\go_{\gl_1}\gd_{t_1t_2}-\go_{\gl_1}^2\callD^{++}_{\gl_1}\(t_1-t_2\)+\\
-\int \di t_3 
\gD\calgP^L_{\gl_1}\(t_1-t_3\)
\callD^{++}_{\gl_1}\(t_3-t_2\),
}
and
\ml{
\frac{\di^2}{\di t_2^2}\callD^{++}_{\gl_1}\(t_1-t_2\)=
-\go_{\gl_1}\gd_{t_1t_2}-\go_{\gl_1}^2\callD^{++}_{\gl_1}\(t_1-t_2\)+\\
-\int \di t_3 
\callD^{++}_{\gl_1}\(t_1-t_3\)
\gD\calgP^R_{\gl_1}\(t_3-t_2\).
}
}
As it will be clear in the next section \e{eq:reg.10} and \e{eq:reg.9} are equivalent when the exact left and right self--energies are used. But they can lead to
different results when the self--energy is approximated.

\section{Screening, double--counting and over--screening}\label{sec:SCR}
%==============================================================================================================================================================
Although the right and left self--energies have  a different analytic structure it is instructive to see why and how
their perturbative expansion coincides. To this end the diagrammatic expansion provides an intuitive and graphical
interpretation. 

The inverse electronic dielectric function $\epsilon^{-1}$ is defined in \e{eq:vertex.4} in terms of the reducible response function, $\chi\(1,2\)$.
The most common approximation for $\chi$ is the Hartree or RPA approximation which  corresponds to assume $\ti{\gC}\(12,3\)\approx \gd_{12}\gd_{13}$.
Within the RPA approximation $\chi(1,2)$ is then written as
\begin{widetext}
\eql{eq:SCR.1}{
  \chi\(1,2\) = \chi^0\(1,2\) +  \int \di 34 \chi^0\(1,3\)v\(3,4\)\chi\(4,2\) =
           \[\chi^0 + \chi^0\otimes v \otimes \chi^0+\dots\]\(1,2\).
}
\begin{figure}[t!]
  {\centering
  \includegraphics[width=\columnwidth]{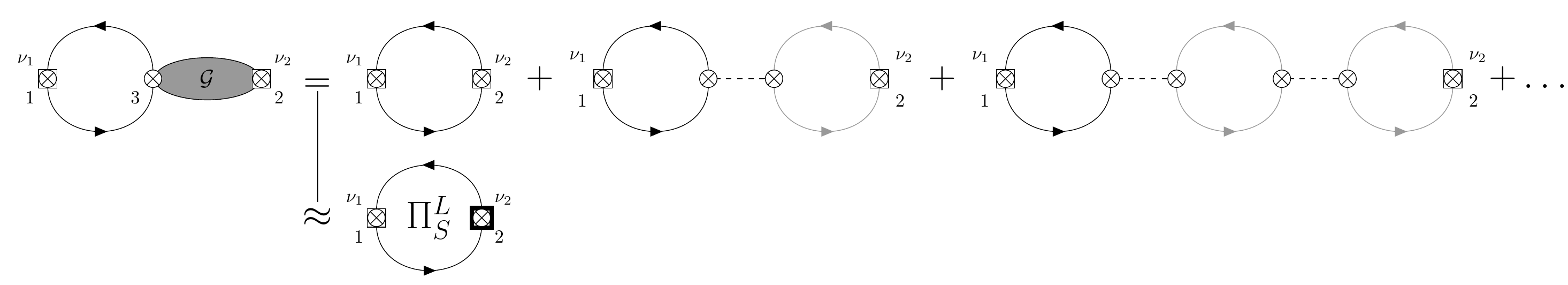}
  }
  {\centering
  \includegraphics[width=\columnwidth]{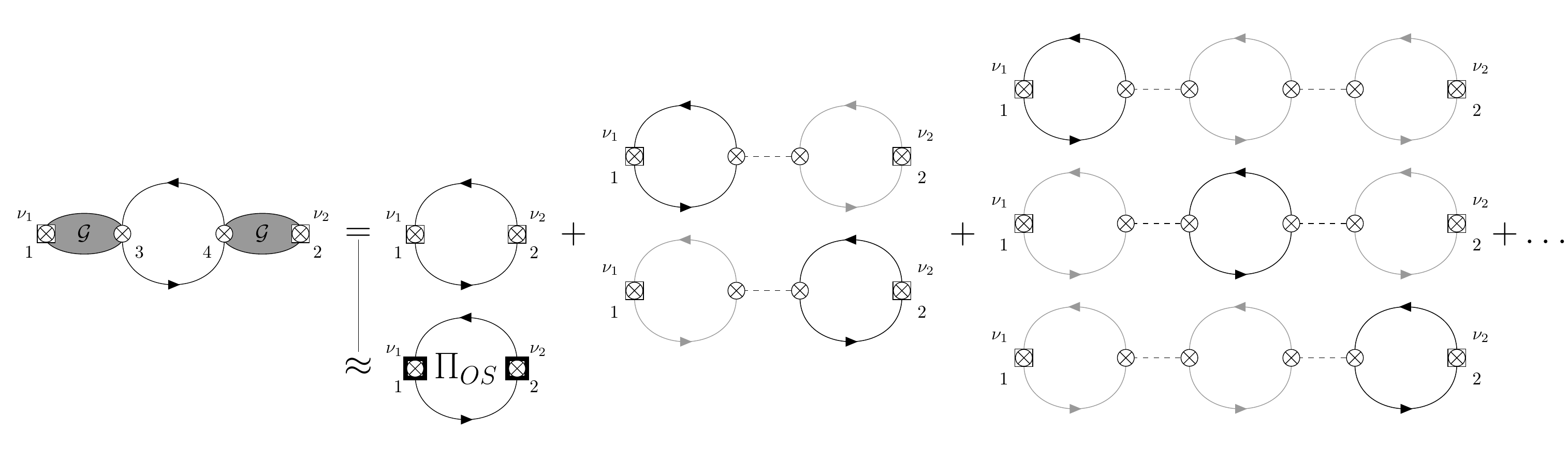}
  }
  \caption{\amend{Diagrammatic representation of the phonon self--energy within the Random--Phase approximation. At the $n$--th order of the perturbative
expansion $\evalat{\calgP^L}{S}$\,(upper frame) contains one diagram, while $\evalat{\calgP}{OS}$\,(lower frame) contains $n+1$ diagrams. The $n$ additional
diagrams in the over--screened case overcount the screening diagrams of the e--p interaction potential.}}
  \label{fig:3}
\end{figure}
\end{widetext}
In the r.h.s. of \e{eq:SCR.1} I have used a compact form ($\otimes$) to represent the spatial convolutions. Within the RPA we have that
\mll{eq:SCR.2}
{
 \evalat{\calgP^L_{\gl_1\gl_2}\(t_1-t_2\)}{RPA}=\\ \int \di\rr_2 \di 3 
g^{\gl_1}\(\rr_1\)
\chi^{0}\(1,3\)
\gee^{-1}\(3,2\)
\callG^{\gl_2}\(t_2,2\).
}
If we now use \e{eq:SCR.1} to expand in powers of $v$ \e{eq:SCR.2} we get the diagrammatic representation of \fig{fig:3}, upper frame. It is 
clear that at any order of the perturbative expansion we have $\evalat{\calgP^R}{RPA}=\evalat{\calgP^L}{RPA}$.

\subsection{The statically screened eletron--nuclei interaction potential}\label{sec:SS}
%--------------------------------------------------------------------------------------------------------------------------------------------------------------
As it will be clear in \sec{sec:merge} \ai\, calculations can easily provide the statically screened e--p potential. Within the MBPT language this corresponds
to the static $\callG$ potential defined as
\eql{eq:SS.0}
{
 \callG^{\gl}\(1,t_2\)\approx \callG^{\gl}_S\(\rr_1\)=\gd_{t_1t_2} \int \di \rr_2 \gee^{-1}\(\rr_1,\rr_2\) g^{\gl_1}\(\rr_2\).
}
If $\gee^{-1}$ is calculated by means of TD--DFT it is possible to \amend{approximate} $\callG^{\gl}_S\(\rr_1\)\approx\callG^{\gl}_{DFPT}\(\rr_1\)$ (see \sec{sec:merge}).

We see immediately, however, that if we apply \e{eq:SS.0} to \e{eq:reg.9} and \amend{to} \e{eq:reg.10} we obtain different results. This is due to the fact that 
if
\seq{
\lab{eq:SS.3}
\eq
{
 \evalat{\calgP^{L}_{\gl_1\gl_2}\(t_1-t_2\)}{S}=\int \di\rr_1 \di \rr_2  \callG^{\gl_1}_S\(\rr_1\)\ti{\chi}\(1,2\)g^{\gl_2}\(\rr_2\),
}
and
\eq{
 \evalat{\calgP^{R}_{\gl_1\gl_2}\(t_1-t_2\)}{S}=\int \di\rr_1 \di \rr_2 g^{\gl_1}_S\(\rr_1\)\ti{\chi}\(1,2\)\callG^{\gl_2}_S\(\rr_2\),
}}
it easily follows that
\eql{eq:SS.4}
{
 \evalat{\calgP^{L}_{\gl_1\gl_2}\(t_1-t_2\)}{S}\neq \evalat{\calgP^{R}_{\gl_1\gl_2}\(t_1-t_2\)}{S}.
}
More importantly \e{eq:SS.3} defines self--energies that are not symmetric under $t_1\lrar t_2$. This breakdown of the time--inversion symmetry is 
inconsistent with the equilibrium regime where the dynamics is invariant under a fixed time translation. As a consequence, for example,
$\evalat{\calgP^{L/R}_{\gl_1\gl_2}\(t_1-t_2\)}{S}$ does not respect the Fluctuation--Dissipation Theorem\,(FDT)~\cite{Leeuwen2013} that, at the equilibrium,
reads
\eql{eq:FDT}
{
 \calgP^{L/R}_{\gl_1\gl_2}\(\go\)=-\frac{1}{\pi}\int d\go^\prime \frac{ \Im\[\calgP^{L/R}_{\gl_1\gl_2}\(\go^\prime\)\]}{\go+i 0^+ -\go^\prime}.
}
\e{eq:FDT} allows, for example, to identify the phonon widths with the $\Im\[\calgP\]$.  This implies that $\evalat{\calgP^{L/R}_{\gl_1\gl_2}\(t_1-t_2\)}{S}$
are nonphysical and cannot be used.

\amend{\e{eq:reg.9}, instead, provides a symmetric form of the self--energy suitable to take the static limit:}
\mll{eq:SS.5}
{
 \evalat{\calgP_{\gl_1\gl_2}\(t_1-t_2\)}{S}=\frac{1}{2}\[\evalat{\calgP^{L}_{\gl_1\gl_2}\(t_1-t_2\)}{S}+\nl+\evalat{\calgP^{R}_{\gl_1\gl_2}\(t_1-t_2\)}{S}\].
}
\e{eq:SS.5} is symmetric under $t_1\lrar t_2$ and respect the FDT, \e{eq:FDT}.

As explained in the \sec{sec:intro.review} several works, instead of using \e{eq:SS.5} has adopted an {\em over--screened}\,(OS) approximation
\eql{eq:SS.6}
{
 \evalat{\calgP_{\gl_1\gl_2}\(t_1-t_2\)}{OS}= \int \di\rr_1 \di \rr_2  \callG^{\gl_1}_S\(\rr_1\) \ti{\chi}\(1,2\) \callG_S^{\gl_2}\(\rr_2\).
}
\e{eq:SS.6} {\em is not} compliant  with the Hamiltonian \e{eq:H.11}. Indeed it can be formally derived only by assuming
\eqgl{eq:SS.7}
{
 \wh{H}_{e-e}=0,\\
 g^{\gl}\(\rr\)\Rar \callG_S^{\gl}\(\rr\).
}
From \e{eq:SS.6} and \e{eq:SS.7} \amend{follow} a series of observations:
\begin{itemize}
\item[i.] 
\elab{eq:SS.7}{a} is not consistent with the original Hamiltonian and, consequently, the Hamiltonian that 
produces $\evalat{\calgP}{OS}$ does not correspond to a physical Taylor expansion of an \ai\, Hamiltonian. 
\item[ii.] \e{eq:SS.6} is not consistent with the adiabatic limit, defined by the reference $\Pi^{ref}$, \e{eq:ref.2}. In $\Pi^{ref}$ only one potential,
$V_{e-n}$ is screened, in agreement with \e{eq:SS.5}.
\item[iii.]
The diagrammatic expansion of $\evalat{\calgP^{L}}{S}$ and $\evalat{\calgP}{OS}$ are shown in \fig{fig:3} within the RPA approximation. In the upper frame
$\calgP^{L}$ is expanded in powers of $v$. We see that at each order of the expansion there is only one contribution. The gray fermionic lines come from the expansion 
of $\callG$. In the lower frame, instead, the same expansion is done for $\evalat{\calgP}{OS}$ and it appears that at the order $n$ of the expansion the OS
self--energy has $n$ equivalent diagrams instead of 1. This means that $\evalat{\calgP}{OS}$ is affected by a severe over--counting of diagrams that, physically,
corresponds to an over--screening of the e--p effective potential.
\end{itemize}

\subsection{Numerical approximations and the \ai\, implementation}\label{sec:theo_implementation}
%--------------------------------------------------------------------------------------------------------------------------------------------------------------
In this section I introduce a simplified form of the phonon self--energy that will be implemented in \yambo\amend{\,(\app{sec:code})} to calculate the phonon
line--widths in an exactly solvable model\,(\sec{sec:jell}) and in a paradigmatic material\,(\sec{sec:mgb2}):
\begin{itemize}
\item[i.] I take the {\em Matsubara} component of the Keldysh expressions with time arguments lying on the imaginary axis. 
\item[ii.] I separate the generalized phonon and electron indices into
branch/band index and momentum index: $\alpha \rightarrow \lambda \qq$, $i \rightarrow n\kk$, $j\rightarrow m\kk^\p$.
\end{itemize}
Being at equilibrium, I switch to frequency space via Fourier transform, considering 
$\calgP_{\gl \qq}\(\im \omega_n\)$. Here  $\omega_n=(2n+1)\pi/\beta$, with $n$ integer and
$\beta$ the inverse temperature, is the Matsubara imaginary frequency.  

After this steps \e{eq:reg.9} equation reduces to 
\mll{eq:eq_limit.1}
{
 \callD^{++}_{\gl \qq}\(\im \omega_n\) = \evalat{\callD^{++}_{\gl \qq}\(\im \omega_n\)}{0}+\\
 \frac{1}{2}\[\evalat{\callD^{++}_{\gl \qq}\(\im \omega_n\)}{0} \gD\calgP^{L}_{\lambda \qq}\(\im \omega_n\) \callD^{++}_{\gl \qq}\(\im \omega_n\)+\nl+
 \callD^{++}_{\gl \qq}\(\im \omega_n\) \gD\calgP^{R}_{\lambda \qq}\(\im \omega_n\) \evalat{\callD^{++}_{\gl \qq}\(\im \omega_n\)}{0}\].
}
where the free $\callD^{++}$ is
\eql{eq:eq_limit.2}{
\amend{\evalat{\callD^{++}_{\lambda \qq}\(\im\go\)}{0}} = -\frac{\gO_{\gl \qq}}{\(\go^2+\gO_{\lambda \qq}^2\)}.
} 
From \e{eq:eq_limit.1} it easily follows that we can define a final  Matsubara self--energy as
\eql{eq:eq_limit.3}{
 \amend{\calgP_{\lambda \qq}\(\im\go\)} = \frac{1}{2}\[ \calgP^{L}_{\lambda \qq}\(\im \omega_n\) + \calgP^{R}_{\lambda \qq}\(\im \omega_n\)\],
} 
so that \e{eq:eq_limit.1} can be finally rewritten as
\eql{eq:eq_limit.4}
{
 \callD^{++}_{\gl \qq}\(\im \omega_n\) = \evalat{\callD^{++}_{\gl \qq}\(\im \omega_n\)}{0}\[ 1 + \gD\calgP_{\lambda \qq}\(\im \omega_n\) \callD^{++}_{\gl \qq}\(\im
\omega_n\)\].
}
Finally, I perform the Matsubara summation of the internal frequency, so that the only integration left is the one over momenta $\kk$.  Furthermore, this latter integration is
discretized as $\int \di^3 \kk/\Omega_{BZ} \rightarrow \sum_k/N_k$.  Here, $\Omega_{BZ}$ is the reciprocal-space volume of the Brillouin zone (BZ), while $N_k$ is
the number of $\kk$-points in a discrete mesh spanning the BZ itself.  
Within the RPA approximation we get 
\eql{eq:eq_limit.5}
{
\evalat{\calgP_{\gl \qq}\(\im\omega_n\)}{kind} = 
\frac{2}{N_k}\sum_{nm\kk}\evalat{\callG^{\lambda \qq}_{mn\kk}}{kind} \frac{f_{m\kk-\qq}-f_{n\kk}}{\im \omega_n +\epsilon_{m\kk-\qq}-\epsilon_{n\kk}},
}
with $kind=\(S,OS\)$.  In \e{eq:eq_limit.3} $\varepsilon_{n\kk}$ and $\varepsilon_{m\kk-\qq}$ are electronic energies, the functions $f_{m\kk-\qq}$ and $f_{n\kk}$ are the temperature-dependent electronic
Fermi-Dirac occupation factors and the prefactor of $2$ comes from the spin summation.  We see from these equations that the \amend{over--screened} and
screened self--energies differ by the coupling strengths $\callG$:
\seq{
\lab{eq:eq_limit.6}
\eq{
 \evalat{\callG^{\lambda \qq}_{mn\kk}}{OS} = |\bra{n\kk}  \callG^{\gl \qq}_S\(\rr\) \ket{m\kk-\qq}|^2,
}
and
\ml{
 \evalat{\callG^{\lambda \qq}_{mn\kk}}{S} = \\ \frac{1}{2} \[ \bra{n\kk}  \callG^{\gl \qq}_S\(\rr\) \ket{m\kk-\qq} \bra{m\kk-\qq}  g^{\gl \qq}\(\rr\)
\ket{n\kk}+\nl
 \bra{n\kk}  g^{\gl \qq}_S\(\rr\) \ket{m\kk-\qq}  \bra{m\kk-\qq}  \callG^{\gl \qq}\(\rr\) \ket{n\kk} \].
}
}

\subsection{Comparison of the screened and over--screened self--energies in an exactly solvable model}\label{sec:jell}
%--------------------------------------------------------------------------------------------------------------------------------------------------------------
In \sec{sec:SCR} I have discussed the analytic properties of the $\evalat{\calgP}{OS}$ self--energy and I have demonstrated that it is not a real MBPT
self--energy. Nevertheless, in \ocite{Calandra2010}, the authors stated that the over--screened approximation better accounts for the error
induced by the static approximation. 

In order to provide further information and solid justifications of the two approximations\,(OS vs S) I consider here
a model e--p Hamiltonian characterized by a single phonon with energy $\go_0$ interacting with a gas of free electrons via a Fr\"{o}hlich like, $q$--dependent
e--p interaction $g_q$. In addition to the e--p term I include the Hartree potential so to describe the dynamical \amend{screening of $g_q$}.
The model Hamiltonian describing this system is
\seq{
\lab{eq:jell.1}
\amend{
\ml{
 \h{H}_m=\sum_\kk \gee_\kk \h{c}^\dagger_{\kk} \h{c}_\kk+\\
  \sum_\qq\[ \go_0\sum_s \(\h{\phi}^\dagger_{s\qq} \h{\phi}_{s\qq} \)+
  \sqrt{2} g_q  \h{\phi}_{+\qq} \gD\h{\gr}_\qq\]+\\
 \frac{1}{\gO}\sum_{\qq}\frac{4\pi}{q^2}\average{\h{\gr}_{-\qq}}\h{\gr}_\qq.
}
with
\eq
{
 \h{\gr}_\qq= \frac{1}{ \sqrt{N}}\sum_{\kk}\h{c}^\dag_{\kk}\h{c}_{\kk-\qq}.
}}
Following \ocite{Nery2018}, I define
\eq
{
 g^2_q=\frac{\ga}{q^2}\frac{2\pi\go_0}{\gO}\sqrt{\(\frac{2\go_0}{m^*}\)},
}
with $\ga$ the a--dimensional e--p Fr\"{o}hlich constant, $q=|\qq|$ and $k=|\kk|$.
}

The energy levels are assumed to be  $\gee_\kk=\frac{\kk^2}{2 m^*}$ with $m^*$ the effective mass. 

As $\h{H}_m$ contains just the Hartree interaction term the exact \amend{real axis phonon self--energy, obtained by 
evaluating \e{eq:eq_limit.5} at $\im \go_n\rar \go+\im 0^+$, is}
\amend{
\eql{eq:jell.3}
{
 \calgP_q\(\go\)=\frac{2}{\gO_{RL}}g_q^2\evalat{\chi_q\(\go\)}{RPA}=\frac{2}{2\gO_{RL}}g_q\chi^0_q\(\go\)\callG_q\(\go\).
}
\elab{eq:jell.3} can be calculated exactly in terms of the independent particle response function $\chi^0$:
}
\seq{
\lab{eq:jell.4}
\eq
{
 \chi^0_q\(\go\)=\int d\kk \frac{f_{\kk-\qq}-f_\kk}{\go+i 0^+ + \gee_{\kk-\qq}-\gee_\kk},
}
and
\eq
{
 \callG_q\(\go\)=g_q \gee^{-1}_q\(\go\).
}
}
In the present case the $S$ and $OS$ self--energies, \e{eq:eq_limit.5}, are
\eqgl{eq:jell.5}
{
 \evalat{\calgP_{q}\(\go\)}{S}=\frac{2}{\gO_{RL}}g_q\chi^0_q\(\go\)\callG_q\(0\),\\
 \evalat{\calgP_{q}\(\go\)}{OS}=\frac{2}{\gO_{RL}}\callG_q\(0\)\chi^0_q\(\go\)\callG_q\(0\).
}

The exact $\chi^0_q\(\go\)$ can be calculated analytically. The mathematical procedure is described in \app{sec:jell.Xo} and the final result is
\ml{
 \label{eq:M.8}
 \Im\[\chi^0_q\(\go\)\]=-\frac{\pi^2\gO m^*}{q}\\
 \sum_{s=\pm 1}\[\gt\(k_F-\left|\frac{\(s\go-\gee_\qq\)m^*}{q}\right|\) \nl
 \(k^2_F-\(\frac{\(s\go-\gee_\qq\)m^*}{q}\)^2\)
 \].
}
The real part of $\chi^0_q\(\go\)$ can be calculated by using the FDT, \e{eq:FDT}, applied to the response function
\eq{
 \label{eq:M.9}
 \chi^0_q\(\go\)=-\frac{1}{\pi}\int d\go^\prime \frac{ \Im\[\chi^0_q\(\go^\prime\)\]}{\go+i 0^+ -\go^\prime}.
}

\e{eq:jell.3} and \e{eq:jell.5} provide an excellent tool to test the validity of the different approximations. In order, however, to validate the model I 
follow the strategy of finding the values of $\gO, \go_0, \ga$ and $m^*$ the provide the best fit of the dielectric properties of a realistic paradigmatic material,
MgB$_2$.

From \elab{eq:jell.4}{b} it is evident that a key quantity that dictates most of the screening properties is the inverse dielectric function,
$\gee^{-1}_q\(\go\)$. In the model Hamiltonian this is exact \amend{within} the RPA approximation. 

\begin{figure}[t!]
  {\centering
  \includegraphics[width=\columnwidth]{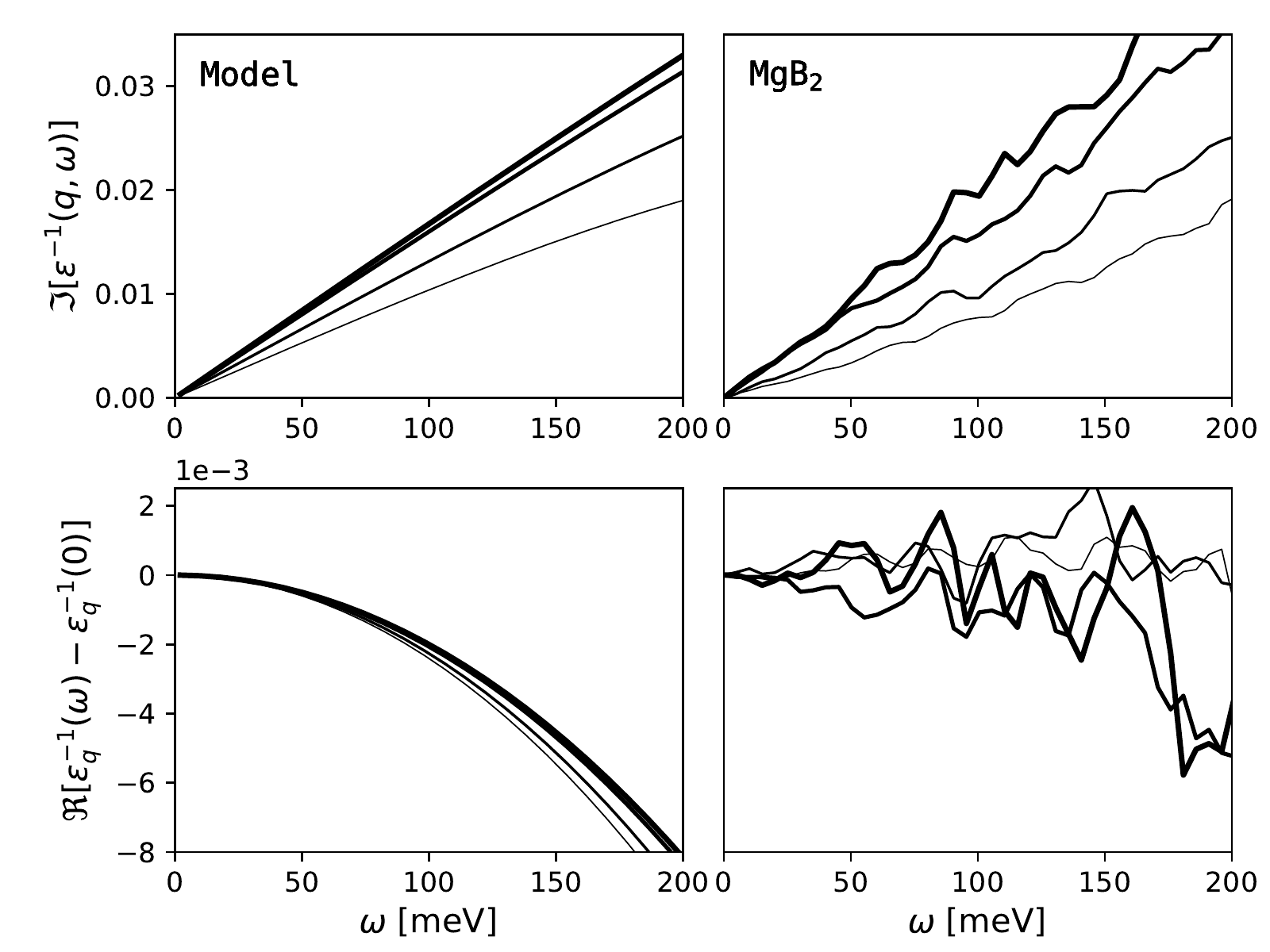}
  }
  \caption{Inverse dielectric function $\gee^{-1}_q\(\go\)$ calculated analytically\,(left frames) and numerically in the case of MgB$_2$\,(right frames) 
  in the low--energy range relevant to the phonon dynamics. 
  The line widths are proportional to $q$ and the left and right frame span the same $\gee^{-1}_q$ range. No re--scaling has been applied.
  The comparison shows that the generalized Fr\"ohlich model Hamiltonian provides an excellent description of the low--energy
  properties of MgB$_2$. 
  }
  \label{fig:4}
\end{figure}
In \fig{fig:4} the $\gee^{-1}_q\(\go\)$  calculated in MgB$_2$ and in the model Hamiltonian are compared for several transferred momenta
and in the energy range $\(0,200\)$\,meV that is relevant for the phonon dynamics. The thickness of the lines is proportional to $q$. From the figure we see that
both the imaginary part and the variation of the real part are well described. In particular the imaginary part shows the same frequency and momentum trend of
the full, \ai\,, calculation. 
The parameters used are tabulated  in \tab{tab:1}.
\begin{table}[h!]
  \begin{center}
    \begin{tabular}{c|c} 
      \textbf{Parameter} & \textbf{Value} \\
      \hline
\amend{$a$} & 20\,a.u. \\
      $\go_0$ & 100\,meV \\
      $\ga$ & 5 \\
      $m^*$ & 0.15 m$_e$ 
    \end{tabular}
  \end{center}
\caption{Generalized Fr\"{o}hlich Hamiltonian parameters used to reproduce the MgB$_2$  $\gee^{-1}_q\(\go\)$. 
\amend{The simulation box is a cubic cell with lattice constant $a$.}}
\label{tab:1}
\end{table}

\begin{figure}[t!]
  {\centering
  \includegraphics[width=\columnwidth]{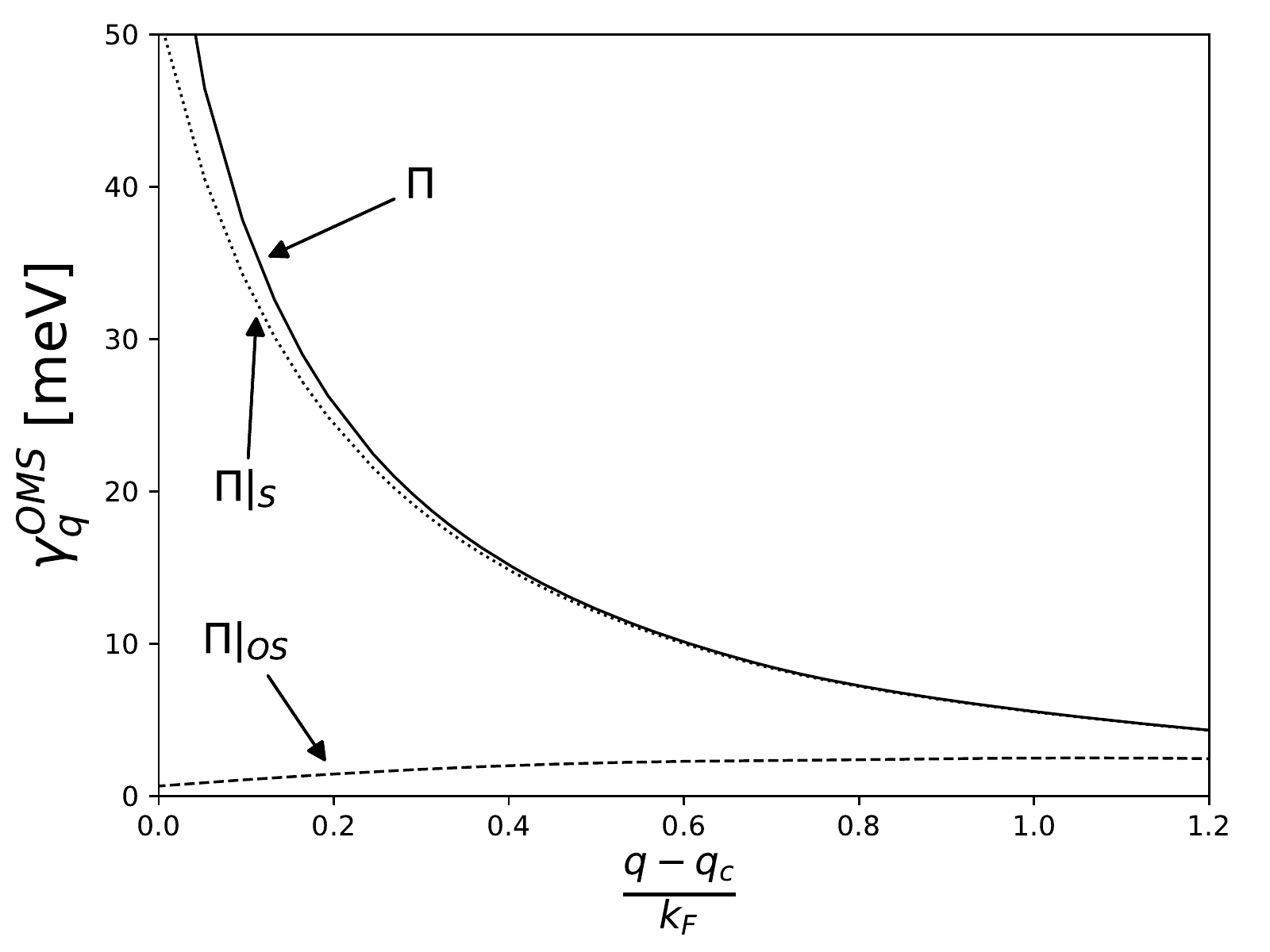}
  }
  \caption{Phonon widths, $\gc_q$ calculated by using the exact self--energy\,(solid), the $\evalat{\calgP}{S}$\,(dotted) and
the $\evalat{\calgP}{OS}$\,(dashed). The over--screened, $OS$, approximation largely underestimates the exact self--energy
and show a wrong long--range, $q\rar 0$, behavior. The momentum range starts from $q_c$ that represents the smallest momentum under which the phonon widths are
zero by definition (see text).
 }
\label{fig:5}
\end{figure}
In \fig{fig:5} I show the $q$ dependency of the on--the--mass shell phonon width, defined as
\amend{
\eql{eq:jell.SS_vs_S.1}
{
 \gc_q= -\Im\[\calgP_q\(\go_0\)\],
}
}
as function of $q-q_c$. $q_c$ is a critical momentum under which the phonon widths are zero by construction. $q_c$ is defined by the 
condition
\eql{eq:jell.SS_vs_S.1.0}
{
 \left|\ s\go_0+\frac{q_c^2}{2m^*}\right|=\frac{k_F q_c}{m^*}\quad\text{with}\, s=\pm 1.
}
$\calgP_q$ is calculated exactly\,(solid line), via \elab{eq:jell.5}{a}\,(dotted line) and via the doubly screened approximation,
\elab{eq:jell.5}{b}\,(dashed line). We see clearly that, while the $\evalat{\calgP_q}{S}$ performs very well for all momentum range, the
$\evalat{\calgP_q}{OS}$ largely underestimates the exact phonon widths. 

\amend{
More importantly $\evalat{\calgP_q}{OS}$ shows a wrong $q\rar 0$ behavior. In order to understand the origin of this we notice that
$\chi^0_q\(\go\) \xrightarrow[q\rar 0]{} O\(1\)$, with the limit taken such that $\go<k_F q$.  From \e{eq:jell.3} it follows that
\eql{eq:jell.SS_vs_S.2}
{
 \callG_q\(0\)=\frac{q^2g_q}{q^2-4\pi\chi^0_q\(0\)} \xrightarrow[q\rar 0]{} q.
}
If we now notice that from \e{eq:M.8} it follows that $ \Im\[\chi^0_q\(\go\)\] \xrightarrow[q\rar 0]{} \frac{1}{q}$ we finally obtain that }
\eql{eq:jell.SS_vs_S.3}
{
 \calgP_q\(\go\)  \xrightarrow[q\rar 0]{} 
 \begin{cases}
  \amend{1/q} & \quad\text{exact}\\
  \amend{1/q} & \quad\text{S}\\
  \amend{q} & \quad\text{OS}
 \end{cases},
}
\amend{and, indeed, in \fig{fig:5} $\evalat{\gc_q}{OS}  \xrightarrow[q\rar 0]{} q$.}

\e{eq:jell.SS_vs_S.3} and \fig{fig:5} \amend{represent} a clear demonstration that $\evalat{\calgP}{OS}$ {\it is not} a Many--Body compliant approximation  and it
leads to a, potentially severe, underestimation of the phonon widths.

\section{On the merging of MBPT with the classical  Born--Oppenheimer approximation}\label{sec:merge}
%==============================================================================================================================================================
In \sec{sec:ref} I have introduced the reference system of phonons without specifying the corresponding atomic coordinates. 
In this section I will discuss the connection between the classical treatment of   \e{eq:H.1} and
MBPT. In particular, in order to bridge the quantistic treatment with DFPT, it is essential to connect 
the different potentials that appear in  \e{eq:H.11} with their classical counterparts.

\begin{figure}[t!]
  {\centering 
  \includegraphics[width=\columnwidth]{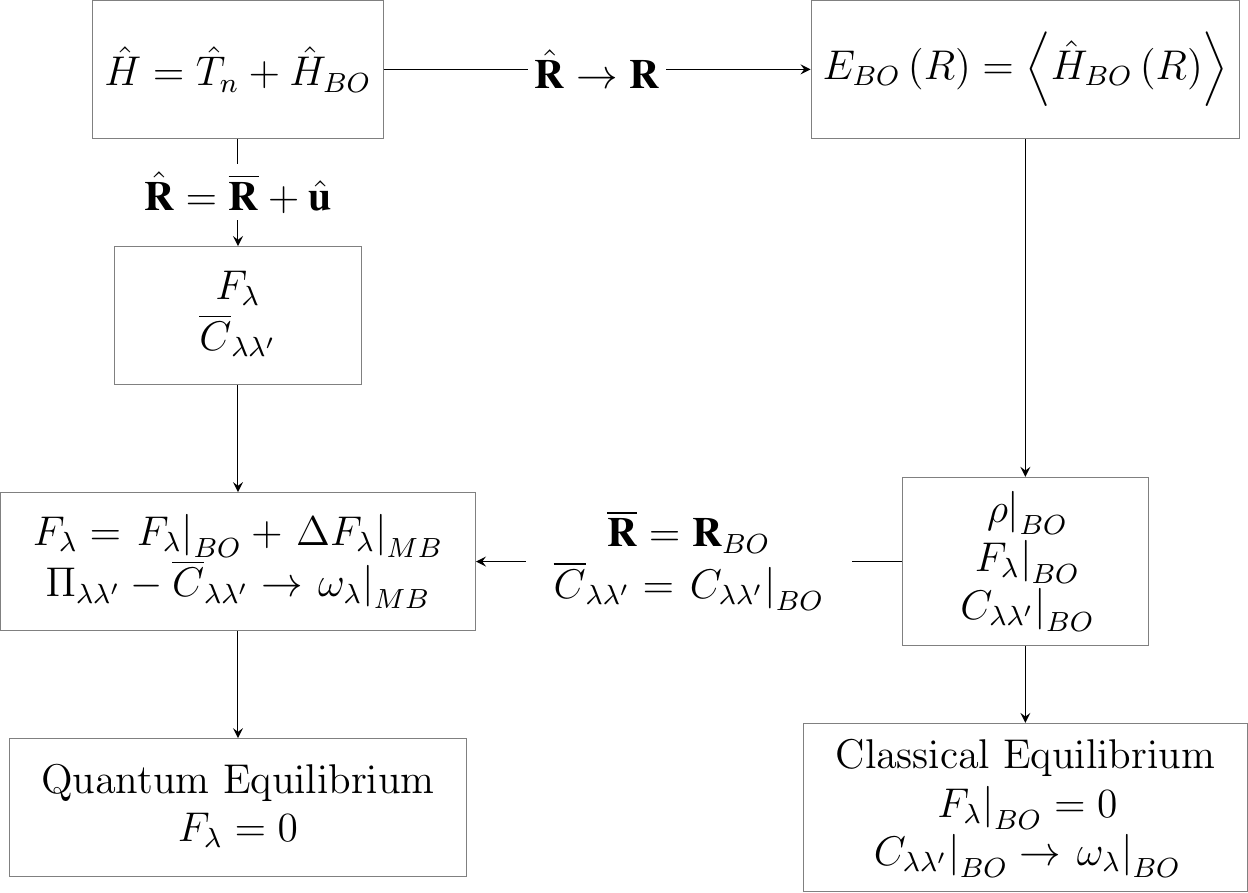}
  }
  \caption{Schematic representation of the connection between the classical harmonic expansion around an arbitrary set of
atomic positions on the BO surface and MBPT.
  \label{fig:6}}
\end{figure}

At this point it is essential to note that the reference BO energy surface, defines only the
reference dynamical matrix, \e{eq:ref.2}. The residual atomic force, \e{eq:ref.1} is defined by the reference
positions. In order to connect those two quantities we need to formally introduce the BO surface
of \e{eq:H.1}. This is obtained by calculating the average of $\h{H}$ without including the 
nuclear kinetic operator and treating the atomic position operators as classical variables:
\eqgl{eq:merge.1}
{
 \h{H}_{BO}=\h{H}_e+\h{H}_{n-n}+\h{H}_{e-n},\\
 E_{BO}\(\RR\)=H_{n-n}\(\RR\) + \average{ \h{H}_e+\h{H}_{e-n}\(\RR\)}.
}
If we now select a specific set of atomic positions, $\RR_{BO}$, we can calculate the corresponding density, $\evalat{\gr\(\rr\)}{BO}$, dynamical matrix 
\eql{eq:merge.2}
{
 \evalat{\t{C}_{IJ}}{BO}=\evalat{\grad_I\grad_J  E_{BO}\(\RR\)}{\RR=\RR_{BO}},
}
and the force $F_I$ 
\eql{eq:merge.3}
{
 \evalat{\FF_{I}}{BO}=-\evalat{\grad_I E_{BO}\(\RR\)}{\RR=\RR_{BO}}.
}
For a generic atomic configuration the force can be non zero \amend{and, as we are not in the equilibrium
configuration, the phonon frequencies can even be negative. At this point we can also define the classical equilibrium condition such that
$\evalat{\FF_{I}}{BO}={\bf 0}$.}
These conceptual steps are schematically described in \fig{fig:6}.

As described at length in this work the fully quantistic treatment of $\h{H}$ implies the presence, in the e--p interaction, of a force term defined by
\e{eq:ref.1}. We can now connect the reference dynamical matrix and the BO by assuming
\eql{eq:merge.4}
{
 \oo{\RR}=\RR_{BO}.
}
It follows that
\eqgl{eq:merge.5}
{
 F^{ref}_\gl=\evalat{F_{\gl}}{BO}+\evalat{\gD F_{\gl}}{MB} ,\\
 C^{ref}_{\gl\gl^\p}=\evalat{C_{\gl\gl^\p}}{BO}.
}
\e{eq:merge.5} is written in the BO phonon basis that, thanks to \e{eq:merge.4}, corresponds to the reference phonon basis. In \elab{eq:merge.5}{a}
\eql{eq:merge.6}
{
\evalat{\gD F_{\gl}}{MB}=\sum_I\int\di\rr \oo{\partial_\gl V_{e-n}\(\rr,\RR_I\)} \[\evalat{\gr\(\rr\)}{BO}-\gr\(\rr\)\]
}
\e{eq:merge.6} defines the MBPT equivalent of the classical equilibrium as the atomic configuration such that the {\em total} force is zero.

The physical picture is that if we start from a BO atomic position the MBPT will feel a force that will be stronger when the initial reference position is far
from a BO equilibrium or there is a strong e--p interaction. In any case, even if we start from a zero classical force configuration, there will still be a residual force that is needed 
to move the system in the MBPT equilibrium. 

It is important to note, however, that if $\gr\(\rr\)\sim \evalat{\gr\(\rr\)}{BO}$ the BO phonons will be renormalized but the equilibrium positions will not
be affected.

\subsection{The case of the Density--Functional Theory reference system}\label{sec:merge.dfpt}
%--------------------------------------------------------------------------------------------------------------------------------------------------------------
Within DFPT atoms are treated classically and the theory is based on the BO.  In practice Density--Functional methods are \ai\, approaches to calculate $E_{BO}\(\RR\)$,
the atomic configuration corresponding to the equilibrium and the oscillations around this equilibrium in a fully self--consistent way and treating correlation
exactly~\cite{R.M.Dreizler1990}.

DFT and DFPT, therefore, are natural candidates to be used as reference system. In this case, by definition, 
\eql{eq:merge.10}
{
\RR_{BO}=\RR_{DFPT}\Rar \evalat{F_{\gl}}{BO}=0,
}
and the MB correction will, eventually, move the equilibrium positions in the new configuration where also $\evalat{\gD F_{\gl}}{MB}=0$.

In addition $\t{C}^{ref}_{IJ}=\evalat{\t{C}^{ref}_{IJ}}{DFPT}$, which implies that \e{eq:ref.2} reduces to
\eql{eq:merge.11}
{
 C^{ref}_{\gl_1\gl_2}= \sum_I\int\di\rr \oo{\partial_{\gl_1} V_{e-n}\(\rr,\RR_I\)}\, \oo{\evalat{\partial_{\gl_2}\gr\(\rr\)}{DFPT}}.
}
\e{eq:merge.11} is a by--product of any DFPT calculation.  It is interesting, however, to investigate if it is possible, and under
which conditions, to connect $C^{ref}$ to the static limit of $\calgP$, defined in \e{eq:SS.5}.

The answer to this question is greatly simplified by the simple form of the exact phonon self--energy, \e{eq:SCR.2}, that is written 
in terms of the irreducible response function and dielectric function. Those two quantities can be indeed calculated by using
DFT. 
The response function defined in \elab{eq:vertex.9}{b} includes e--e and e--n correlation
effects. Indeed also the inverse dielectric depends on the atomic fluctuations. In the electronic case this is a well known effect that 
leads to the non--diagonal Debye--Waller correction~\cite{Gonze2011}. 

If we neglect e--n effects in the response function and dielectric function we can approximate the mass operator with the the DFT exchange--correlation potential:
\eql{eq:merge.13}
{
 M\(1,2\)\approx \gd_{12}V_{xc}\(1\).
}
Thanks to \e{eq:merge.13} we can calculate exactly all ingredients of $\calgP$ by using DFT and it follows that, at the equilibrium,
\amend{$\calgP_{\gl_1\gl_2}\(\go\)$ reduces} to 
\begin{widetext}
\amend{
\eql{eq:merge.14}
{
 \calgP_{\gl_1\gl_2}\(\go\) \xrightarrow[\go\rar 0]{DFT} \int \di  \rr_1 \di \rr_2 g^{\gl_1}\(\rr_1\)
\evalat{\chi_0\(\rr_1,\rr_2;\go=0\)}{DFT}\evalat{\callG^{\gl_1}\(\rr_2\)}{DFT}.
}
We can now notice that, thanks to Kubo
\eql{eq:merge.15}
{
 \int \di \rr_2 \evalat{\chi_0\(\rr_1,\rr_2;\go=0\)}{DFT}\evalat{\callG^{\gl_2}\(\rr_2\)}{DFT}=\oo{\evalat{\partial_{\gl_2}\gr\(\rr_1\)}{DFPT}},
}}
\end{widetext}
which leads to the final result
\eql{eq:merge.16}
{
 \evalat{\calgP_{\gl_1\gl_2}\(\go\)}{S} \xrightarrow[\go\rar 0]{DFT}  C^{ref}_{\gl_1\gl_2}.
}
\e{eq:merge.16} means that if we use a mean--field, DFT approximation for the electronic self--energy and, in addition we take the static limit of 
all components of the phonon self--energy we obtain that the static phonon self--energy coincides with the static electron--nuclei component of the DFPT density matrix.

Another consequence of using DFT to describe the electronic linear--response  is that we can link the static limit of the
time--dependent effective e--n interaction potential, \e{eq:vertex.7}, \amend{to} the DFPT potential
\eql{eq:merge.17}
{
 \evalat{\callG^{\gl}\(\rr_1\)}{DFPT}=\int \di \rr_2 \gee^{-1}_{DFT}\(\rr_1,\rr_2\) g^{\gl_1}\(\rr_2\).
}
$\evalat{\callG^{\gl}}{DFPT}$ is encoded in several public \ai\, codes and it is the by--product of any DFPT phonon calculation.

At this point it is crucial to observe that \e{eq:merge.16} does  not hold for the over-screened self--energy:
\begin{widetext}
\eql{eq:merge.18}
{
 \evalat{\calgP_{\gl_1\gl_2}\(\go\)}{OS} \xrightarrow[\go\rar 0]{DFT} \int \di  \rr_1 \di \rr_2 \evalat{\callG^{\gl_1}\(\rr_1\)}{DFT}
 \evalat{\chi_0\(\rr_1,\rr_2;\go\)}{DFT}\evalat{\callG^{\gl_1}\(\rr_2\)}{DFT}\neq C^{ref}_{\gl_1\gl_2}.
}
\end{widetext}

\section{Results in a paradigmatic material: MgB$_2$}\lab{sec:mgb2}
%==============================================================================================================================================================
\begin{figure}[t!] 
 {\centering 
 \includegraphics[width=\columnwidth]{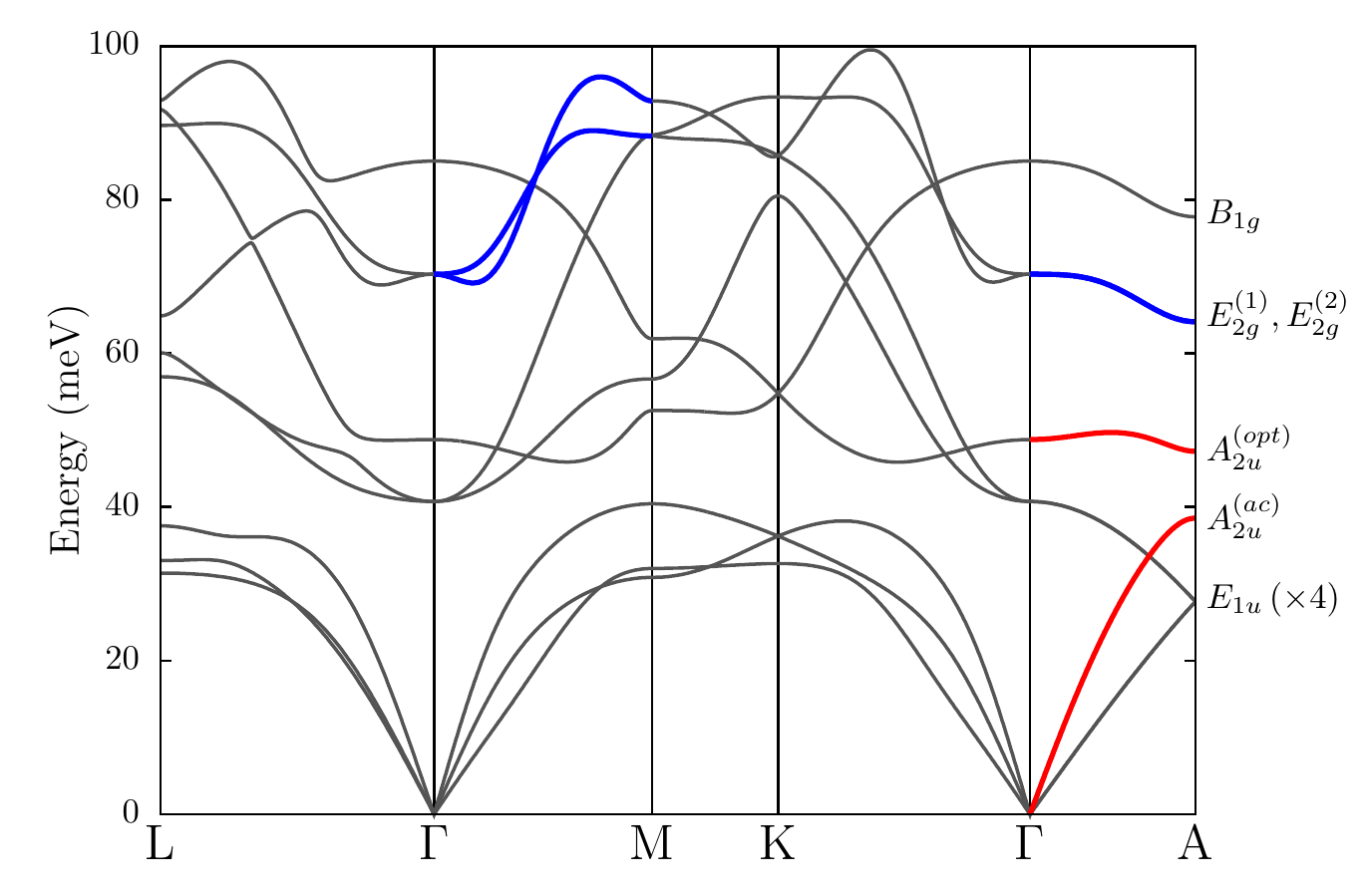}
 } 
 \caption{Phonon band structure of MgB$_2$. The phonon mode symmetries at the $A$ point are given. The modes and $\qq$-regions exhibiting large coupling to the
 $\sigma$ and $\pi$ bands are emphasized in blue and red, respectively, and are analyzed in \fig{fig:9}.
\label{fig:7}} 
\end{figure}

Magnesium diboride, MgB$_2$ is a metallic layered material composed of alternating $2$D sheets of boron and magnesium.
It transitions to phonon--mediated superconductivity at the critical temperature $T_c= 39$ K\cite{Nagamatsu_2001}.  This behavior is almost entirely
due to e--p coupling relative to the boron atoms, whose electrons form in--plane $\sigma$ and out--of--plane $\pi$ bonds.  These bonds are in turn responsible for the
existence of two superconducting band gaps, with different theoretical $T_c$\cite{Choi_2002_Nature,Margine_2013,Floris_2005}.  I will briefly discuss the
connection between phonon widths and super--conductivity in \sec{sec:closing.SC}. 
What is relevant in the present context is that the super--conductive properties of MgB$_2$ clearly point to a strong e--p coupling. 
In particular, the $\sigma$ bands are considered to yield a giant ``anomalous'' e--p coupling due to the strong orbital overlap induced by the
in-plane optical phonon mode $E_{2g}$, as opposed to the $\pi$ bands undergoing a weak coupling.  The calculated phonon dispersion are reported in \fig{fig:7}
and the numerical details, code developments and calculation flow are discussed in \app{sec:code}.

The $E_{2g}$ phonon linewidths have been extensively studied, both theoretically and experimentally, along the
$\Gamma$A\cite{Shukla2003,Calandra2005,Calandra_2007} and $\Gamma$M\cite{Baron_2004} directions in the hexagonal Brillouin zone.  
In these studies, the comparison is made between the full--width half--maximum of inelastic x--ray scattering spectral peaks
and the phonon widths. As mentioned in the introduction all calculations performed so far 
have used the over--screened formulation, \e{eq:SS.6}.
The results show in general a reasonable agreement, though particularly along $\Gamma$A the experimental linewidths are found to be larger than the theoretical
results.  For example, in \ocite{Shukla2003} a theoretical value of $20.35$ meV is found at point A, while the experimental peak width is closer to $30$
meV.

\begin{figure}[t!]
  {\centering 
  \includegraphics[width=\columnwidth]{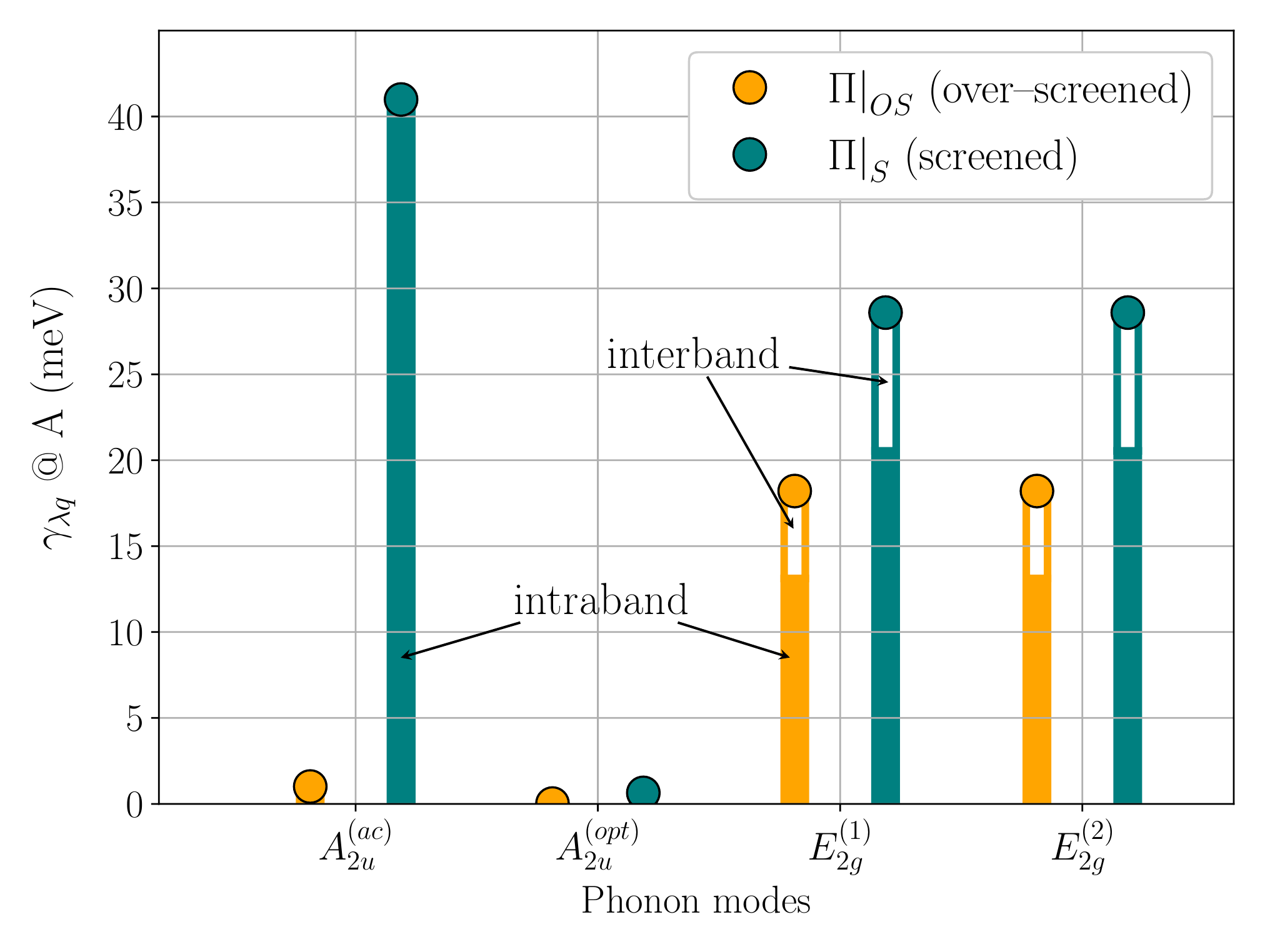}
  }
  \caption{Phonon linewidths in MgB$_2$ computed at the A point in the Brillouin zone. Only the relevant phonon modes are shown. Orange: over--screened
($\calgP_{OS}$) linewidths. Teal: screened ($\calgP_{S}$) linedwiths  The full and empty bars represent the contributions to the linewidths stemming from intraband and
interband processes, respectively.\label{fig:8}}
\end{figure}

\begin{figure}[t!]
  {\centering
  \includegraphics[width=\columnwidth]{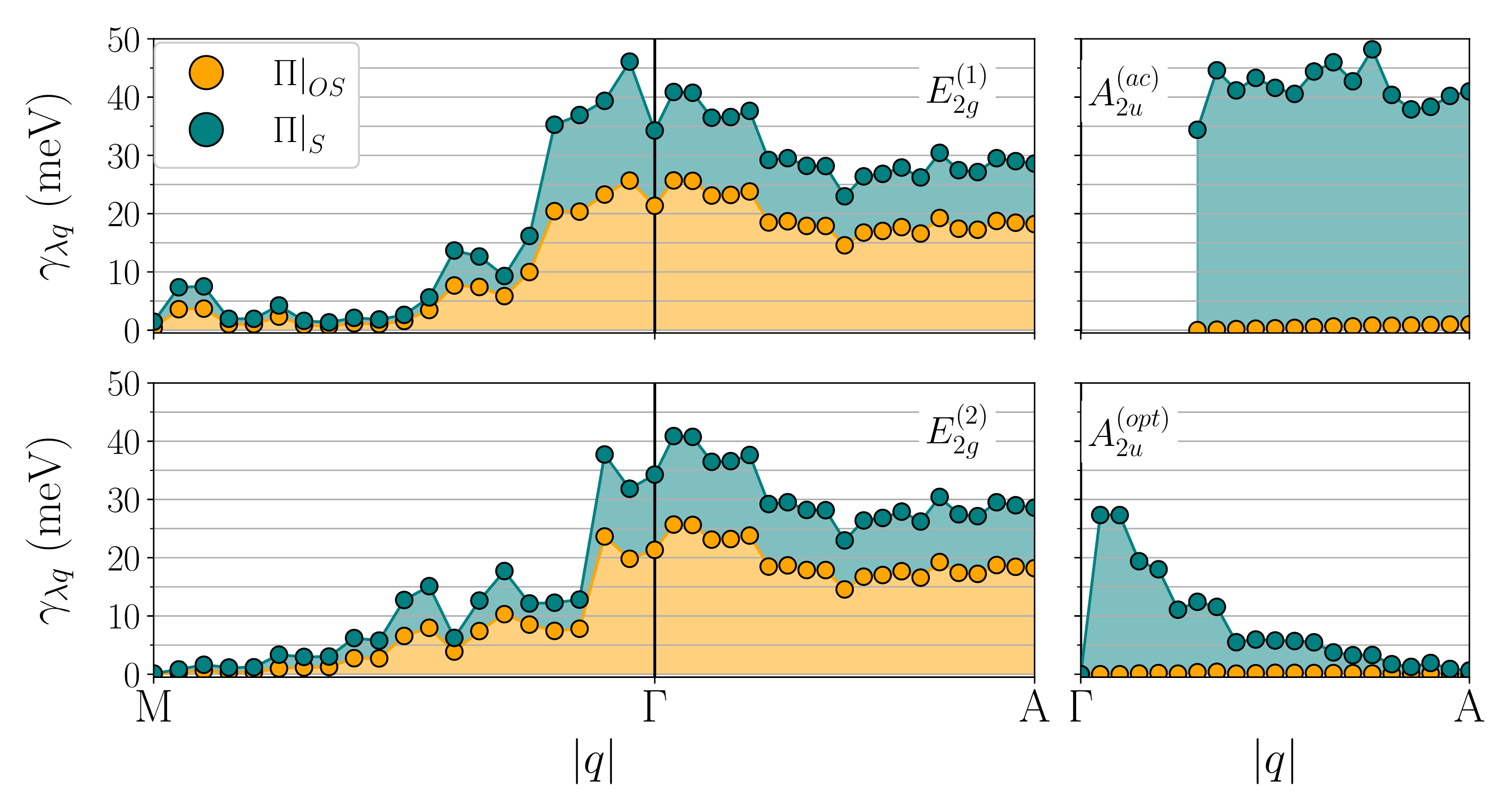}
  }
  \caption{Phonon linewidths of (a) the $E_{2g}$ optical modes along the M$\Gamma$A path and (b) the $A^{\(ac\)}_{2u}$ acoustic (top) and optical,
$A^{\(opt\)}_{2u}$ (bottom), modes along $\Gamma$A. Orange: over-screened (OE) coupling. Teal: screened (no OE) coupling.\label{fig:9}}
\end{figure}

Let me start the discussion from the calculation of the $E_{2g}$ mode.  As can be seen from \fig{fig:8}, I obtain $18.2$ meV in the over-screened case, 
in very good agreement with the same calculation in \ocite{Shukla2003}.  The screened case, \e{eq:SS.5}, gives instead the
value of $28.6$ meV, showing a $57\,\%$ increment in the phonon linewidths.  Let's notice that in the case of $\sigma$ bands, around $70 \%$ of the contribution
comes from ``intraband'' terms (i.e., elecron-hole pairs are formed within the same $\sigma$ subband), while the remaining $30 \%$ is due to ``interband'' terms
involving different $\sigma$ subbands.  The large $57 \%$ increase in the linewidths is also the average along the full $\Gamma$A direction, as can be seen from
\figlab{fig:9}{a}, while a strong increase also appears along the $\Gamma$M direction.  Along the latter, both over-screened and screened linewidths sensibly decrease
after the midpoint from $\Gamma$ to M due to a sharp increase of the relative phonon energies.  The comparison with experiment is difficult due to the large
error-bars, but overall we do obtain a better agreement compared to  the over-screening case (compare with Fig. 3 in \cite{Shukla2003} and Fig. 3 in \cite{Baron_2004} for
the $\Gamma$A and $\Gamma$M directions, respectively).

The $E_{2g}$ mode is not the only one undergoing large changes when over-screening is removed.  In fact, we see from \fig{fig:8} that also the acoustic
$A^{\(ac\)}_{2u}$ mode gains a giant linewidth increase from $1$ meV (over-screened case) to $41$ meV (screened case).  By looking at \fig{fig:9}(b) we realize that these giant
linewidths appear along the full $\Gamma$A direction, where the acoustic $A^{\(ac\)}_{2u}$ mode maintains an average linewidth of $42$ meV, and also characterize the higher-energy,
optical mode of the same symmetry. 
Now, the infrared-active $A_{2u}$ modes involve out-of-plane oscillations of the boron atoms; furthermore, we see that the linewidth is composed by purely
intraband, jellium--like, contributions. As discussed in \sec{sec:jell} the OS self--energy largely underestimates the phonon widths practically for
most of the $|\qq|$ values. This underestimation contributes to  the large enhancing of the widths in the  $\Gamma$A direction of MgB$_2$.
Our results suggest that the $\pi$-band e--p coupling may also be anomalous.

Interestingly, the acoustic $A_{2u}$ linewidths remain constant along $\Gamma$A despite strong variations in the relative phonon energies, while the optical
$A_{2u}$ linewidths decrease from $28$ to $0.6$ meV despite the phonon energies being roughly constant.  We also point out that such a giant over-screening
effect is not limited to the coupling with the $\pi$ bands along $\Gamma$A, but also appears -- although to a lesser extent -- in the linewidths of the in-plane
$E_{1u}$ modes, which couple with the $\sigma$ bands, along $\Gamma$M (here the largest effect is on the higher-energy acoustic $E_{1u}$ mode up to
$0.5\Gamma$M, where the screened linewidths rise to $15-20$ meV).

\section{Closing remarks}\lab{sec:closing}
%==============================================================================================================================================================
Due to the many implications of this work I conclude it by splitting the discussion in three sections. In the first\,(\sec{sec:closing.summary}) I will summarize the theoretical 
implications of using an over--screened formulation for the phonon self--energy. In the second\,(\sec{sec:closing.SC}) I will briefly discuss the implications, mostly
conceptual, on the calculation of the Eliashberg function when it is written in terms of the phonon widths. I will then, finally, conclude with a general section of
conclusions in \sec{sec:conclusions}.

\subsubsection{Summary of the evidences of an over--screening  error in the phonon self--energy}\lab{sec:closing.summary}
%--------------------------------------------------------------------------------------------------------------------------------------------------------------
One of the goals of this work is to propose a controllable approximation to  the phonon self--energy, $\calgP$ that can be merged with DFPT making possible accurate \ai\,
calculations of phonon properties.

\begin{table}[h!]
  \begin{center}
    \begin{tabular}{c|c|c|c|c} 
      \textbf{Self--energy} & \textbf{MBPT} & $q\rar 0$ & $D_n$ & $\go\rar 0$ \\
                            & \textbf{compliant} &  &  &  $C^{ref}_{\gl_1\gl_2}$\\
                            & Sec.\ref{sec:SCR} & Sec.\ref{sec:jell} & Sec.\ref{sec:SS} & Sec.\ref{sec:merge.dfpt}  \\
      \hline
      $\calgP$ & yes & \amend{ $1/q$} & 1 & yes \\
      $\evalat{\calgP}{S}$ & yes & \amend{ $1/q$} & 1 & yes \\
      $\evalat{\calgP}{OS}$ & no & \amend{$q$} & n+1 & no 
    \end{tabular}
  \end{center}
\caption{Summary of the properties of the exact phonon self--energy compared to the screened and over--screened approximations.}
\label{tab:2}
\end{table}

From a general point of view any approximation to $\calgP$ should be {\it controllable}. For {\it controllable} I mean that the error induced by the use of specific 
approximations should be estimated, even if roughly, and connected to characteristic physical properties of the materials.  
A \amend{family} of properties that can help in defining a {\it controllable} approximation are the exact limits, $\qq\rar 0$ and $\go \rar 0$.

In the present case I have used the following conditions to define a {\it controllable} approximation to $\calgP$
(also tabulated in \tab{tab:2}):

{\it MBPT compliant}.
The series of approximations used must be applied to an initial formulation that, being MBPT, admits a diagrammatic expansion. To be MBPT compliant means
to respect the basic rules of the diagrammatic expansion like the Fluctuation--Dissipation theorem and the time reflection symmetry. 
This point was studied in detail in \sec{sec:SCR} where I showed that the over screened approximation
over--counts bubble diagrams and, therefore, is not MBPT compliant. In practice this means that if $D_n$ is the number of bubbles at the $n$--th order of the
diagrammatic expansion the over--screened approximation has $n$ diagrams more compared to the screened and exact expressions\,(\sec{sec:SS}).

{\it $q\rar 0$ and $\go\rar 0$}. Two very important limits that must be respected by the approximation are the static and zero momentum limits. As discussed in
\sec{sec:merge.dfpt} the over screened approximation does not reduce to the reference, adiabatic, dynamical matrix  that is written in terms of a singly screened e--p
potential.  At the same time the exact solution of the generalized Fr\"ohlich Hamiltonian, discussed in \sec{sec:jell}, has demonstrated that 
the  over screened approximation has a wrong $q\rar 0$ limit.

The conclusion of this section is, thus, that the over--screened approximation, affected by the over--screening error, is not a {\it controllable} and
physically sound MBPT approximation.

\subsubsection{Implications on the calculation of the Eliashberg function}\lab{sec:closing.SC}
%--------------------------------------------------------------------------------------------------------------------------------------------------------------
In 1972, P.B.Allen~\cite{Allen1972}, introduced a formulation of the Eliashberg function where 
\eql{eq:conc.1}{
\ga^2F\(\go\)\propto\sum_{\qq \gl} \gc_{\gl \qq}\gd\(\go-\gO_{\gl\qq}\).
}
The basic idea of the work of Allen was to make possible to calculate the spectral function by using 
quantities accessible in experimental, neutron scattering, experiments.

As already noted by P.B.Allen in 1983~\cite{ALLEN19831} and demonstrated here the phonon widths defined in \e{eq:conc.1} are affected by a severe
over--screened error. It is important to note that the expression of the Eliashberg function in terms of the e--p potential is well--known and correct. 
The over--screening effect appears if the phonon widths are connected to $\ga^2F\(\go\)$ via \e{eq:conc.1}.

\subsubsection{Conclusions}\lab{sec:conclusions}
%--------------------------------------------------------------------------------------------------------------------------------------------------------------
In this work I have re--analyzed the Many--Body description of the phonon dynamics from several aspects.  I have developed a general framework to evaluate the
phonon self--energy that admits a controllable static screening approximation avoiding the over--screening error. 

By reviewing and extending the literature I demonstrate that 
the inclusion of all \ai\, e--n and n--n potentials leads to additional force and quadratic terms in the  e--p Hamiltonian. These novel terms are shown to 
be essential in the merging of MBPT with \ai\, Density--Functional theories.

The equilibrium Dyson equation for the phonon displacement Green's function has been derived on the Keldysh contour and the equilibrium limit has been carefully
derived. I showed that there exists three equivalent \amend{formulations} in the equilibrium limit of which only one is suitable to take the static screening
approximation.  This formulation allows \amend{for} a formal static limit without breaking the time reflection symmetry, needed for obtain a controllable and 
Many--Body  compliant approximation.

The final, symmetric, expression of the phonon self--energy has been compared with the exact solution of a generalized Fr\"ohlioch Hamiltonian. I showed that the
over--screened approximation fails in describing the $\qq$ exact dependence of the \amend{$\calgP_q\(\go\)$}  that, instead, is well described by the screened
approximation proposed here.

I also provided a first--principles numerical scheme for the calculation of over--screening error--free phonon linewidths in the equilibrium case, at no
additional cost with respect to the state-of-the-art, systematically over--screened approach.  This can be applied to any system whose mean--field description is
accessible via DFPT.  The scheme is applied to MgB$_2$ where I demonstrate several important implications of using the proposed screened 
phonon self--energy.
 
The final results of this work lead to implications in many applications where the phonon dynamics plays a crucial role. This ranges from phonon widths and
energies renormalizations due to non--adiabatic effects to real--time processes as involved in thermal transport and lattice dynamics.

\section{Acknowledgments}
%==============================================================================================================================================================
A.M. gratefully acknowledges: 
Andrea Recchia, Jan Berges, Dino Novko for discussions about the technical aspects of the phonon widths calculations;
Fulvio Paleari for the priceless numerical support in the coding and evaluation of the MgB$_2$ results;
Enrico Perfetto and Gianluca Stefanucci for the enlightening discussions of the subtle theoretical aspects hidden in the derivation. Their work 
on the electron--boson ultrafast dynamics~\cite{PhysRevLett.127.036402} has given me the idea of deriving the equilibrium limit down-folding the first--order
Baym--Kadanoff equations.
A.M. acknowledges the funding received from the European Union projects: MaX {\em Materials design at the eXascale} H2020-INFRAEDI-2018-2020/H2020-INFRAEDI-2018-1, Grant agreement n. 824143;  
{\em Nanoscience Foundries and Fine Analysis -- Europe | PILOT}  H2020-INFRAIA-03-2020, Grant agreement n. 101007417; 
{\em PRIN: Progetti di Ricerca di rilevante interesse Nazionale} Bando 2020, Prot. 2020JZ5N9M.

\appendix

\section{Explicit expression for the $\callD^{s_1s_2}_{\gl_1 \gl_2}\(t_1-t_2\)$ equations of motion}\label{sec:D_matrix_eom}
%==============================================================================================================================================================
The equation of motion for the components of the equilibrium $\ul{\callD}$ can be found by 
expanding the r.h.s. of \es{eq:pi.1}{eq:pi.1.a}. For the left derivative we have
\seq{
\lab{eq:reg.3}
\ml
{
\frac{\di}{\di t_1} \callD^{--}_{\gl_1 \gl_2}\(t_1-t_2\)=-\go_{\gl_1} \callD^{+-}_{\gl_1\gl_2}\(t_1-t_2\)-\\
\int \di t_3 \sum_{\gl_3} \gD\calgP^L_{\gl_1\gl_3}\(t_1-t_3\) \callD^{+-}_{\gl_3 \gl_2}\(t_3-t_2\),
}
\ml
{
\frac{\di}{\di t_1} \callD^{-+}_{\gl_1 \gl_2}\(t_1-t_2\)=-\go_{\gl_1} \callD^{++}_{\gl_1\gl_2}\(t_1-t_2\)-\\
\int \di t_3 \sum_{\gl_3} \gD\calgP^L_{\gl_1\gl_3}\(t_1-t_3\) \callD^{++}_{\gl_3 \gl_2}\(t_3,t_2\)-\gd_{t_1t_2},
}
\eq
{
\frac{\di}{\di t_1} \callD^{+-}_{\gl_1 \gl_2}\(t_1-t_2\)=\go_{\gl_1} \callD^{--}_{\gl_1\gl_2}\(t_1-t_2\)+\gd_{t_1t_2},
}
\eq
{
\frac{\di}{\di t_1} \callD^{++}_{\gl_1 \gl_2}\(t_1-t_2\)=\go_{\gl_1} \callD^{-+}_{\gl_1\gl_2}\(t_1-t_2\).
}
}
In the case of the right derivative it follows that 
\seq{
\lab{eq:reg.4}
\ml
{
\frac{\di}{\di t_2} \callD^{--}_{\gl_1 \gl_2}\(t_1-t_2\)=-\go_{\gl_2} \callD^{-+}_{\gl_1\gl_2}\(t_1-t_2\)-\\
\int \di t_3 
\sum_{\gl_3} 
\callD^{-+}_{\gl_1 \gl_3}\(t_1-t_3\)\gD\calgP^R_{\gl_3\gl_2}\(t_3-t_2\),
}
\eq
{
\frac{\di}{\di t_2} \callD^{-+}_{\gl_1 \gl_2}\(t_1-t_2\)=\go_{\gl_2} \callD^{--}_{\gl_1\gl_2}\(t_1-t_2\)+\gd_{t_1t_2},
}
\ml
{
\frac{\di}{\di t_1} \callD^{+-}_{\gl_1 \gl_2}\(t_1-t_2\)=-\go_{\gl_2} \callD^{++}_{\gl_1\gl_2}\(t_1-t_2\)-\\
\int \di t_3 
\sum_{\gl_3} 
\callD^{++}_{\gl_1 \gl_3}\(t_1-t_3\)\gD\calgP^R_{\gl_3\gl_2}\(t_3-t_2\)+\gd_{t_1t_2},
}
\eq
{
\frac{\di}{\di t_2} \callD^{++}_{\gl_1 \gl_2}\(t_1-t_2\)=\go_{\gl_2} \callD^{+-}_{\gl_1\gl_2}\(t_1-t_2\).
}
}

\section{The exact independent particle response function in the extended Fr\"ohlich Hamiltonian}\label{sec:jell.Xo}
%==============================================================================================================================================================
I now consider the zero--temperature case where $f_\kk=\gt\(\gee_\kk-\frac{k_F^2}{2m^*}\)$. This implies that the 
$\kk$ summation is restricted to $\left|\kk\right|<k_F$.

We now rewrite \elab{eq:jell.4}{a} as
\seq{
\lab{eq:M.2}
\eq{
 \chi^0_q\(\go\)=J_q\(\go\)+J^*_q\(-\go\),
}
with
\eq{
 J_q\(\go\)=\int d\kk f_{\kk}\(\go+i 0^+ + \gee_{\kk}-\gee_{\kk-\qq}\)^{-1}.
}
}
I now notice that if we move to spherical coordinates with the $\h{z}$ axis along $\qq$ we get
\eq{
\lab{eq:M.3}
 J_q\(\go\)=2\pi\int_{-1}^1 dx \int_0^{k_F} k^2 \(\go-\frac{q^2}{2m^*}+\frac{kqx}{m^*}+i0^+\)^{-1}.
}
Let's now focus on the $\Im\[J_q\(\go\)\]$. In \e{eq:M.3} appears $\gd\(\go-\frac{q^2}{2m^*}+\frac{kqx}{m^*}\)$ which implies
\eq{
\lab{eq:M.4}
 x=-\frac{\(\go-\frac{q^2}{2m^*}\)m^*}{kq}.
}
We now distinguish two cases: $\go=\frac{q^2}{2m^*}$ and $\go\neq\frac{q^2}{2m^*}$.

{\em Case I: $\go=\frac{q^2}{2m^*}$}.
In this case the integral in \e{eq:M.3} is straightforward and gives:
\eq{
\lab{eq:M.5}
 \evalat{\Im\[J_q\(\go\)\]}{\go=\frac{q^2}{2m^*}}=-\frac{\pi^2 m^* k_F^2}{q}.
}

{\em Case II: $\go\neq\frac{q^2}{2m^*}$}.
In this case the $k$ range in \e{eq:M.3}  is $k\in\[k_0,k_F\]$ with
\eq{
\lab{eq:M.6}
 k_0=\left|\go-\frac{q^2}{2m^*}\right|\frac{m^*}{q}.
}
The integral in \e{eq:M.3} reduces to
\eq{
\lab{eq:M.7}
 \evalat{\Im\[J_q\(\go\)\]}{\go\neq\frac{q^2}{2m^*}}=-\frac{\pi^2 m^*}{q}\(k_F^2-k_0^2\)\gt\(k_F-k_0\).
}
By using \e{eq:M.5} and \e{eq:M.7} in \e{eq:M.2} we finally obtain \e{eq:M.8}.

\section{Code development}\label{sec:code}
%==============================================================================================================================================================
The codes used for the implementation of this work and the subsequent numerical calculations were Quantum Espresso (QE)\cite{QE-2017} for the DFT and DFPT
steps, and Yambo\cite{YAMBO-2019} for the calculation of the phonon linewidths.  Below I discuss the code implementation that was necessary to compute the
equilibrium phonon self--energy, and next the general scheme of a linewidth calculation.

\subsection{Quantum Espresso}\lab{sec:code_imp.QE}
%--------------------------------------------------------------------------------------------------------------------------------------------------------------
The bare electron-phonon matrix elements, $g^{\lambda q}_{mnk}$,  \elab{eq:H.9}{a}, were extracted from a QE-DFPT calculation by modifying the part
relative to the \texttt{ph.x} executable and in particular the subroutines contained in \texttt{/PHonon/PH/elphon.f90} so that the bare matrix elements could be
stored and printed in a format readable by Yambo.  
The modifications were done on version 6.6 of the QE distribution.\footnote{The distribution I used is
available at this address: \url{https://github.com/QEF/q-e/releases/tag/qe-6.6}}. Note that the (complex) spatially integrated matrix
elements of the (real) variation of the bare e--p interaction is directly extracted.

\subsection{\yambo}\lab{sec:code_imp.Y}
%--------------------------------------------------------------------------------------------------------------------------------------------------------------
\es{eq:SS.5}{eq:SS.6} were implemented in \yambo -- version 5.0\footnote{The yambo distribution is
available here \url{https://github.com/yambo-code/yambo/releases/tag/5.0.2}, while the experimental \textit{yambopy} package is here
\url{https://github.com/yambo-code/yambopy}.} -- as part of the ``phonon'' project relative to the \texttt{yambo\_ph} executable.  The implementation is fully
parallel.  

Since \es{eq:SS.5}{eq:SS.6} require a very large $\kk$-point mesh in reciprocal space to be accurately converged, a double grid support
was added in order to compute electronic eigenvalues on a finer grid with respect to the one used for the e--p calculations.  Let me denote the fine grid
as $FG$ and the original, coarse grid as $CG$.  From now on, reciprocal--space points belonging to the $FG$ ($CG$) are written as lowercase $\kk$ (uppercase
$\KK$).  

Let's start rewriting \e{eq:eq_limit.5} for a $\QQ$--point in the CG  as
\eqgl{eq:app.1}
{
\evalat{\calgP_{\gl \QQ}\(\go\)}{kind} = 
 \frac{1}{N_k}\sum_{nm\kk}\evalat{\callG^{\lambda \QQ}_{mn\kk}}{kind} F^{mn\lambda}_{\kk\kk-\QQ}\(\go\),\\
F^{mn\lambda}_{\kk\kk-\QQ}\(\go\)= \frac{f_{m\kk-\QQ}-f_{n\kk}}{\go +\epsilon_{m\kk-\QQ}-\epsilon_{n\kk}+\im 0^+},
}
Now the $\sum_{\kk}$ in \elab{eq:app.1}{a} is represented as a product of sum in the $CG$ and in
the $FG$. The $FG$ can be both a regular or random grid (we used the latter):
\mll{eq:app.2}
{
 \evalat{\calgP_{\gl \QQ}\(\go\)}{kind} = 
 \sum_{\KK\in CG}\sum_{nm}\evalat{\callG^{\lambda \QQ}_{mn\KK}}{kind} \\
  \frac{1}{N_\KK N_{\KK-\QQ}} \sum_{\kk\in FG_{\KK}} \sum_{\pp\in FG_{\KK-\QQ}} F^{mn\lambda}_{\kk \pp}\(\go\)
}
Note here that the $FG$ depends on the $CG$ it was generated from.  In particular,
$N_k$ is the number of $CG$ points in the Brillouin Zone (BZ), while $\sum_{k\in FG_{\KK}}$ represents a sum over the subset of the $FG$
random $k$-points which are closest to each $\KK$-point of the original $CG$.  
The number of $k$-points contained in each $FG_{\KK}$ subset (which
may vary because of randomness and when close to the BZ edge) is $N_\KK$.  For each $\KK$-point, the $FG$ subsets around $\KK$ and
$\KK-\QQ$ are both needed.  

Crucially, both the $CG$ and the $FG$ must undergo convergence tests: a python workflow using the
\textit{yambopy} package was created to automatically generate $CG$-$FG$ pairs.  The numerical evaluation of the delta functions involves the broadening
parameter $\eta$, which has to be chosen -- naturally, as small as possible -- according to the densities of the $CG$ and $FG$ grids.  In addition, in order to
avoid the unnecessary, time- and memory-expensive counting of transitions contributing negligibly, Yambo automatically selects only transitions satisfying
$\varepsilon_{n\kk}-\varepsilon_{m\kk-\qq}<=\Omega_{\lambda \qq}\pm 3\eta$.

\section{Calculation flow}\lab{sec:code.flow}
%--------------------------------------------------------------------------------------------------------------------------------------------------------------
\begin{figure}[hbtp]
  {\centering 
  \includegraphics[width=\columnwidth]{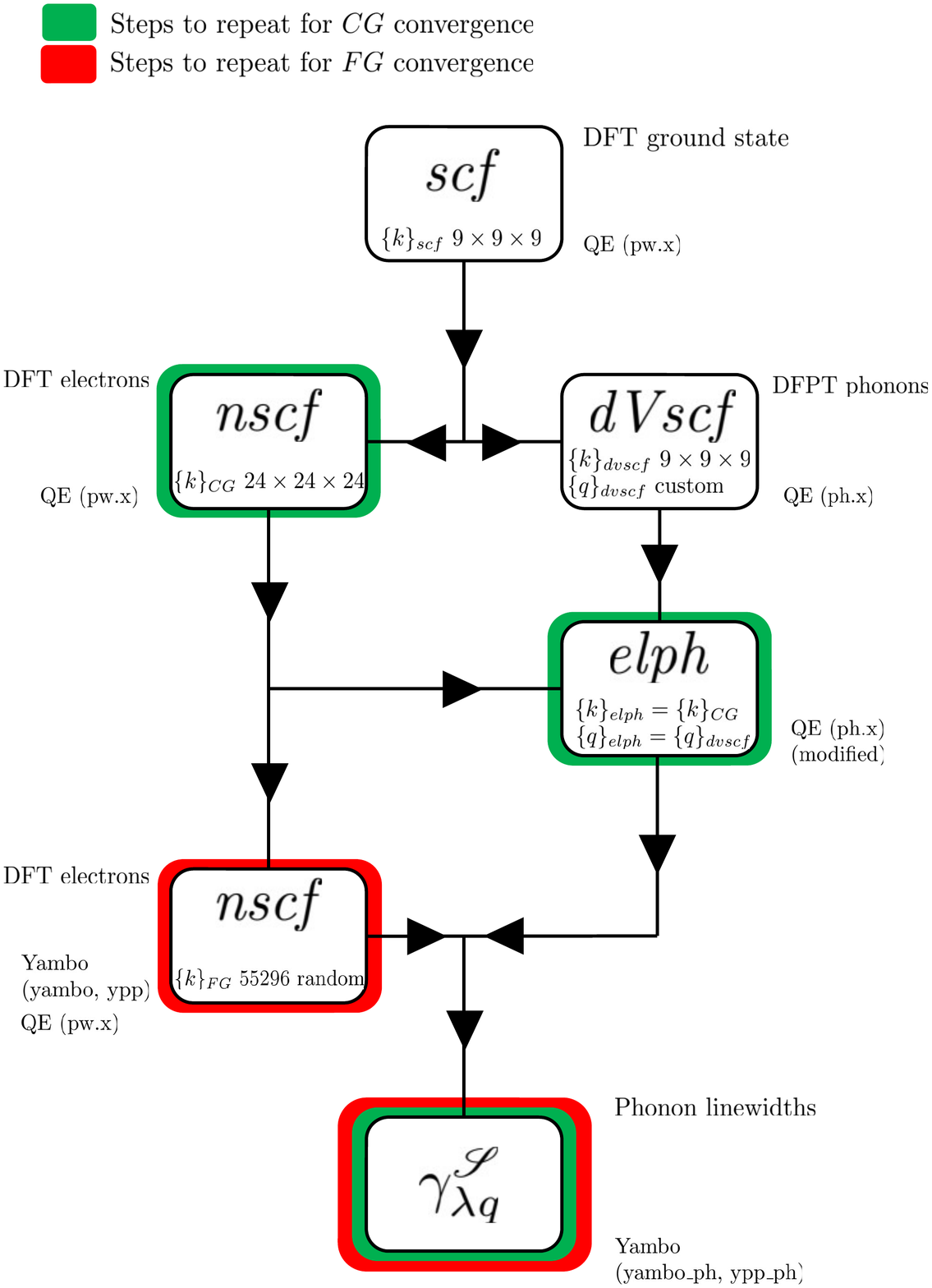}
  } 
  \caption{Linewidths calculation flow discussed in \app{sec:code.flow}. The calculations corresponding to the various steps are shown in boxes. If two
boxes are connected by an arrow, it means that the calculation at the ending point depends on the calculation at the starting point. The various calculations
may depend on different meshes of reciprocal--space points ($\{\kk\}$ for electrons, $\{\qq\}$ for phonons): these meshes are shown in the boxes along with the
values  used in the case of the MgB$_2$ calculations. On the side of the boxes, additional descriptions of the type of calculations are
provided, together with the software package needed (QE or Yambo) and the specific executables (\texttt{pw.x, ph.x, yambo, ypp, yambo\_ph, ypp\_ph}) in
brackets. The green (red) frame denotes the calculations which depend on the \textit{coarse grid} $CG$ of $\KK$-points (\textit{fine grid} $FG$ of
$\kk$-points). See text for more information.\label{fig:10}}
\end{figure}

The \textit{ab initio} phonon linewidths calculation comprises the following six interdependent steps, which are described in detail in \fig{fig:10} and are briefly
summarize here.  
In order to give an assessment of the numerical load, the scheme also lists the converged values of the various reciprocal--space grids used to
calculate the linewidths in MgB$_2$. 

\amend{Besides the $\kk$ and $\qq$--grids an important ingredient of the calculations is the pseudo--potential\,(PP). In this work 
I tested two kind of PP's: a soft PP and a hard PP. In the soft PP the pseudo Boron atom valence comprises the $2p^1$ orbitals, while in the 
hard case the PP includes also the $2s^2$ levels.}

\amend{The hard and soft PP are characterized by two very different wave--functions cutoffs: 1200\,Ry\,(hard PP) and 70\,Ry\,(soft PP). A crucial property of the phonon
self--energy is that if we look at \e{eq:pi.2} we see that, as $C^{ref}_{\gl\gl}$ is a real matrix it follows that it does not contribute to the phonon
widths. Consequently these are more sensible to the details of the calculation, including the kind of PP used.}

\amend{Indeed while both PP's yield the same structural and electronic properties a residual difference remain the case of the phonon widths. In order to
estimate this effect let's consider the $\Gamma\rar A$ direction. If we look at the maximum phonon widths along this line we observe that for the $E^1_{2g}$
state both the screened and over--screened widths change of around 6\% when moving from the soft to the hard PP. 
In the $E^2_{2g}$ case, instead, the screened case suffers a 25\% enhancing, while the over--screened case changes only of the 6\%. Finally, while
the $A^{ac}_{2u}$ state remains unchanged for both the screened and over--screened widths the screened $A^{opt}_{2u}$ state is quenched when the hard PP is
used.}
\amend{The final result is that the message of this work is not at all affected by the use of a soft or hard PP. Still a quantitative evaluation of the
phonon widths requires a more careful investigation of the role played by the core levels and the pseudo--potential approximation}.

\amend{Once grids and PP's are selected here it follows} the calculation flow:
\begin{itemize}
\item[(i)] Self-consistent-field (scf) ground-state calculation using a regular $\{\kk\}_{scf}$ grid.
\item[(ii)] Derivatives of the scf potential (dVscf) and interatomic force constants calculation using a regular $\{\kk\}_{dvscf}$ grid. This fixes the list of
phonon momenta $\qq$, which may be automatically generated (regular grid) or properly chosen.
\item[(iii)] Non-self-consistent-field (nscf) calculation. The $CG$ grid used in this calculation defines the $\KK$-points and has to be
carefully converged together with the $FG$ grid from Step (v) and the broadening parameter $\eta$. The grids convergence can be tested on the
phonon linewidths calculations $\gamma_{\lambda \qq}$ -- step (vi) -- looking both at $\qq$-averages along the BZ and at the values at high-symmetry
$q$-points. This step also computes the electron energies $\varepsilon_{n\KK}$.
\item[(iv)] Electron-phonon matrix elements (elph) calculation. In this step -- to be run on top of Steps (ii) and (iii) -- the
$g^{\lambda \qq}_{mn\kk}$ are computed.
\item[(v)] Second non-self-consistent-field (nscf) calculation. This fixes the $FG$, which defines the $\kk$-points. Random $\kk$-points are used since they yield
faster convergence, and the randomly--distributed $FG$  points can be generated with \yambo. The $FG$ has to be carefully converged together with the
$CG$ from Step (iii) and the broadening parameter $\eta$.  It is in this step that the fine-grid electron
energies $\varepsilon_{n\kk}$ are computed.
\item[(vi)] Phonon linewidths calculation. This is the final calculation that yields $\gamma_{\lambda \qq}$. 
\end{itemize}

\FloatBarrier

\bibliography{paper.bib}

\end{document}